\documentclass[a4paper,11pt]{article}

\usepackage{amssymb}  
\usepackage{amsmath}
\usepackage{amsthm}
\usepackage{amstext}
\usepackage{amscd}

\usepackage[margin=1in]{geometry}

\newtheorem{theorem}{Theorem}[section]
\newtheorem{lemma}[theorem]{Lemma}
\newtheorem{corollary}[theorem]{Corollary}

\newtheorem{definition}[theorem]{Definition}

\newcommand*{\textfrac}[2]{{#1}/{#2}}

\newcommand*{\cH}{\mathcal{H}}

\newcommand*{\cN}{\mathcal{N}}

\newcommand*{\cY}{\mathcal{Y}}
\newcommand*{\cZ}{\mathcal{Z}}

\newcommand*{\fY}{\mathfrak{Y}}
\newcommand*{\fZ}{\mathfrak{Z}}

\newcommand*{\bz}{\mathbf{z}}
\newcommand*{\bzb}{\mathbf{\bar{z}}}
\newcommand*{\bzp}{\mathbf{z'}}
\newcommand*{\bZ}{\mathbf{Z}}
\newcommand*{\bZb}{\mathbf{\bar{Z}}}

\newcommand*{\eps}{\varepsilon}
\newcommand*{\epsp}{\bar{\eps}}

\newcommand*{\id}{\mathrm{id}}

\DeclareMathOperator*{\Lin}{End}
\DeclareMathOperator*{\Herm}{Herm}

\DeclareMathOperator*{\ExpE}{\mathbb{E}}
\DeclareMathOperator*{\Exp}{\mathbb{E}}
\DeclareMathOperator*{\PrS}{\mathbb{P}}
\DeclareMathOperator*{\PrE}{\mathbb{P}}
\newcommand*{\Tr}{\mathrm{Tr}}
\newcommand*{\tr}{\mathrm{tr}}
\DeclareMathOperator*{\cdist}{\delta}
\DeclareMathOperator*{\qdist}{\delta}

\newcommand*{\ket}[1]{| #1 \rangle}
\newcommand*{\bra}[1]{\langle #1 |}
\newcommand*{\spr}[2]{\langle #1 | #2 \rangle}

\newcommand*{\conc}[2]{({#1},{#2})} 

\newcommand*{\Hyp}[3]{\operatorname{Hyp}(#1, #2, #3)}
\newcommand*{\POVM}[1]{\operatorname{POVM}_{#1}}

\newcommand*{\expdist}[2]{D_{#2}(#1)} 
\newcommand*{\notimes}{} 
\newcommand*{\notimesn}{\!} 
\newcommand*{\freq}[1]{Q_{#1}}
\newcommand*{\freqfix}{q} 
\newcommand*{\freqfixp}{\tilde{q}} 

\newcommand*{\matone}[2]{{\hspace{-.17em}\begin{smallmatrix}
 #1 \\[-0.155ex] #2 
\end{smallmatrix}}}
\newcommand*{\mattwo}[4]{{\hspace{-.17em}\begin{smallmatrix}
 #1 & \hspace{-.3em} #2 \\[-0.155ex] #3 & \hspace{-.3em} #4
\end{smallmatrix}}}
\newcommand*{\mattwosmall}[4]{\raisebox{0.1ex}{$\hspace{-.17em}
    \begin{smallmatrix} #1 & \hspace{-.3em} #2 \\[-0.4ex] #3 &
      \hspace{-.3em} #4
\end{smallmatrix}$}}
\newcommand*{\matn}[6]{{\hspace{-.17em}\begin{smallmatrix}
 #1 & \hspace{-.3em} #2 & \hspace{-.3em} \cdots & \hspace{-.3em} #3
 \\[-0.155ex] #4 & \hspace{-.3em} #5 & \hspace{-.3em} \cdots & \hspace{-.3em} #6
\end{smallmatrix}}}

\newcommand*{\tensspace}[1]{{\overline{#1}}}
\newcommand*{\probdistr}{\mathrm{Distr}} 
\newcommand*{\mresultA}[2]{\matone{#1}{#2}}
\newcommand*{\mresultAB}[4]{\mattwo{#1}{#2}{#3}{#4}}
\newcommand*{\states}{\mathrm{S}}

\newcommand*{\partialtrA}[1]{\Tr_{#1}}
\newcommand*{\AcondB}[2]{\mattwosmall{#1}{}{}{|#2}}
\newcommand*{\mresultAcondB}[3]{\mattwo{#1}{}{#2}{|#3}}

\newcommand*{\partAB}[2]{^{{#1}\notimes{#2}}}
\newcommand*{\partA}[1]{^{#1}}
\newcommand*{\Cone}{C_1}
\newcommand*{\Ctwo}{C_2}
\newcommand*{\Cdim}{\bar{C}}

\begin{document}

\title{\bf A de Finetti Representation for Finite Symmetric Quantum
  States\footnote{This work was partially supported by the Swiss
    National Science Foundation, project No.~200020-103847/1.}  }

\author{ Robert K\"onig and Renato Renner\vspace{2ex} \\
             Computer Science Department \\
             ETH Z\"urich;
             Switzerland \\
             {\tt rkoenig@inf.ethz.ch} \quad {\tt renner@inf.ethz.ch}}

\date{}

\maketitle

\abstract{ Consider a {\em symmetric} quantum state on an $n$-fold
  product space, that is, the state is invariant under permutations of
  the $n$ subsystems. We show that, conditioned on the outcomes of an
  informationally complete measurement applied to a number of
  subsystems, the state in the remaining subsystems is close to having
  product form. This immediately generalizes the so-called de Finetti
  representation to the case of finite symmetric quantum states.  }

\section{Introduction}

The analysis of physical experiments is often based on the assumption
that the same experiment can be repeated many times independently. In
particular, one usually assumes that the results $Z_1, \ldots, Z_n$
obtained from $n$ repetitions of the same experiment are distributed
according to some product distribution, i.e., $P_{Z_1 \cdots Z_n} =
(P_Z)^n$. In practical situations, however, the independence of the
individual outcomes $Z_i$ can usually not be guaranteed.

The so-called de Finetti representation theorem~\cite{deFinetti37} can
be seen as a solution to this problem\footnote{See~\cite{deFinetti93}
  for a collection of de Finetti's original papers.}. Basically, it
states that the assumption on the product structure of $P_{Z_1 \cdots
  Z_n}$ can be replaced by a seemingly weaker assumption, namely that
the distribution of the outcomes of infinitely many repetitions of the
experiment are invariant under reordering. For instance, this is the
case if the $n$ samples $Z_1, \ldots, Z_n$ are randomly chosen from
infinitely many repetitions of the experiment.

Let us briefly explain this result on a more formal level. We say that
an $n$-partite probability distribution $P_{Z_1 \cdots Z_n}$ is
\emph{symmetric} if it is invariant under any permutation of the
random variables $Z_1, \ldots, Z_n$. If $P_{Z_1 \cdots Z_n}$ is the
marginal of a symmetric distribution $P_{Z_1 \cdots Z_m}$ over $m \geq
n$ random variables, then $P_{Z_1 \cdots Z_n}$ is called
\emph{$m$-exchangeable}. Moreover, $P_{Z_1 \cdots Z_n}$ is
\emph{infinitely exchangeable} if it is $m$-exchangeable for all $m
\geq n$. The result of de Finetti now states that any infinitely
exchangeable probability distribution $P_{Z_1 \cdots Z_n}$ can be
written as a convex combination of product distributions of the form
$(P_Z)^n$.

This result has been generalized in different directions. Diaconis and
Freedman~\cite{DiaFre80} analyzed the structure of $m$-exchangeable
probability distributions of $n$ random variables, for $n \leq m <
\infty$. This is of particular interest for practical applications,
where the number of experiments is only finite.  They found that, for
appropriate values of $n$ and $m$, these distributions are still close
to the convex hull of the set of product distributions.

The result of de Finetti has also been extended to quantum states, to
which the notion of symmetry and exchangeability can be adapted in an
obvious way. Hudson and Moody~\cite{HudMoo76} showed that any
infinitely exchangeable quantum state $\rho^{\tensspace{n}}$ over $n$
subsystems is a convex combination of product states, i.e.,
$\rho^{\tensspace{n}} = \sum_{\bz} p_{\bz} (\rho_{\bz})^{\otimes n}$,
for appropriate weights $p_{\bz}$ (see also~\cite{Hudson81}). An
alternative proof of this claim has recently been presented by Caves,
Fuchs, and Schack~\cite{cavesfuchsschack} (see
also~\cite{fuchsschacksecond}), relying on the original result of de
Finetti.

In this paper, we analyze the structure of $m$-exchangeable quantum
states over $n$ subsystems, for $n \leq m < \infty$. In a sense, our
result combines the two mentioned directions of generalizing de
Finetti's result.  Note that any $m$-exchangeable state
$\rho^{\tensspace{n}}$ over $n$ subsystems can be extended to an
$m$-exchangeable state $\rho^{\tensspace{n+k}}$ over $n+k$ subsystems,
for $n+k \leq m$. We show that the state $\rho^{\tensspace{n}}_{\bz}$
of the first $n$ subsystems, conditioned on the outcomes $\bz=(z_1,
\ldots, z_k)$ of an informationally complete measurement applied to
each of the remaining $k$ subsystems, is close to a product state
$(\rho_{\bz})^{\otimes n}$. In particular, since
$\rho^{\tensspace{n}}$ can be written as a convex combination of the
states $\rho^{\tensspace{n}}_{\bz}$, i.e., $\rho^{\tensspace{n}} =
\sum_{\bz} p_{\bz} \rho^{\tensspace{n}}_{\bz}$, this immediately
implies that $\rho^{\tensspace{n}}$ is close to the convex combination
$\sum_{\bz} p_{\bz} (\rho_{\bz})^{\otimes n}$.  As in the classical
case, the distance between the $m$-exchangeable state
$\rho^{\tensspace{n}}$ and the convex hull of the set of product
states depends on the values $m$ and $n$. In particular, if $m$ is
much larger than $n$, we obtain the result of~\cite{HudMoo76}
and~\cite{cavesfuchsschack}.

Our result has applications in quantum information theory and, in
particular, quantum cryptography, e.g., for proving the security of
quantum key distribution (QKD) schemes (e.g., \cite{BB84}). An elegant
way to study QKD protocols is to subdivide their analysis into two
conceptually different stages~\cite{Ekert91}.  In the first stage,
Alice and Bob use a quantum channel to distribute $n$ pairs of
entangled particles.\footnote{In many practical QKD protocols, Alice
  prepares particles according to some classical randomness and then
  uses the quantum channel to send them to Bob. However, for analyzing
  such a protocol, one can equivalently think of a protocol where
  Alice prepares pairs of fully entangled particles and then sends one
  particle of each pair to Bob while keeping the other one.} In the
second stage, Alice and Bob each measure their particles and then use
the resulting classical information to generate their keys.

For proving the security of such a QKD scheme, one has to show that,
even if an adversary can manipulate the state $\rho^{A B}$ of the $n$
particle pairs arbitrarily, either the key generated by Alice and Bob
is secure, or they recognize that something went wrong and abort the
protocol.  However, as the state $\rho^{A B}$ can be arbitrary, these
proofs are rather complicated and mostly restricted to a special type
of protocol (see also the discussion in~\cite{ChReEk04}). In this
context, our results allow to reduce the analysis of $\rho^{A B}$ to
the analysis of states which are close to having product
form.\footnote{More precisely, conditioned on the outcome of certain
  statistical tests performed during the protocol (e.g., for
  estimating the error rate), the state of any small subset of
  particle pairs randomly chosen by Alice and Bob is close to a
  product state.} Such product states correspond to the much simpler
situation where the adversary is restricted to attacking each particle
individually.

\subsection*{Outline of the paper}

In Section~\ref{sec:pre}, we introduce some basic notation and
definitions, including the notion of symmetry and exchangeability.
Additionally, we briefly review the properties of the variational
distance between probability distributions, as well as its quantum
analogue, the trace distance between density operators.
Sections~\ref{sec:icpovmandduals}--\ref{sec:mainresult} are devoted to
the proof of our main results on the structure of symmetric quantum
states.  Generally speaking, our proof is based on the analysis of the
statistics obtained when applying informationally complete POVMs to
symmetric quantum states. We will thus be interested in good POVMs in
the sense that the measurement statistics gives maximal information
about the measured state.  Constructing such POVMs is the main purpose
of Section~\ref{sec:icpovmandduals}, which is somehow independent of
the remaining part of the paper.  In
Section~\ref{sec:symmetricprobdistr}, we analyze classical symmetric
probability distributions and derive bounds on their distance to
product distributions.  In Section~\ref{sec:symstates}, it is shown
how to deduce structural properties of quantum states from the
corresponding properties of the measurement statistics, using the
POVMs constructed in Section~\ref{sec:icpovmandduals}. Finally, in
Section~\ref{sec:mainresult}, we combine these results to obtain our
main statements, including a de Finetti representation for finitely
exchangeable quantum states. Additionally, in
Appendix~\ref{sec:markov}, we present an alternative version of our
results which might be more suitable for certain applications.

\section{Preliminaries} \label{sec:pre}

\subsection{Density operators, POVMs, and probability distributions} \label{sec:densop}

Throughout this paper, we will restrict our attention to
finite-dimensional Hilbert spaces, denoted by $\cH$ or $\cH_A$, for
some index $A$. Let $\Lin(\cH)$ be the set of endomorphisms on $\cH$,
and $\Herm(\cH)$ the set of hermitian endomorphisms on $\cH$. An
element $\rho \in \Herm(\cH)$ is called a \emph{density operator} or,
equivalently, a \emph{quantum state} on $\cH$ if it is positive
semidefinite, $\rho \geq 0$, and has trace one, $\tr(\rho) = 1$.  We
denote by $\states(\cH)$ the set of density operators on $\cH$. A
\emph{positive operator valued measure (POVM)} on $\cH$ is a family
$\fZ = \{F_z\}_{z \in \cZ}$ of nonnegative operators $F_z \in
\Herm(\cH)$, $F_z \geq 0$, such that $\sum_{z \in \cZ} F_z =
\id_{\cH}$. The POVM $\fZ$ is called \emph{informationally complete}
if it is a basis of $\Herm(\cH)$.

To improve the readability of formulas involving density operators
on product spaces, we use superscripts to indicate which subsystems an
operator acts on, e.g., we write $\rho^{A B C}$ for a density operator
on $\cH_A \otimes \cH_B \otimes \cH_C$.  Operators with the same name
but different superscripts are related to each other by the partial trace. For
example, $\rho^{A B}$ is the partial state obtained from $\rho^{A B
  C}$ by tracing over $\cH_C$, i.e., $\rho^{A B} =
\partialtrA{C}(\rho^{A B C})$, and, similarly, $\rho^{A} =
\partialtrA{B C}(\rho^{A B C})$.  This notation is consistent since
partial traces over different subsystems commute, e.g., we have
$\rho^{A} = \partialtrA{B}(\rho^{A B}) = \partialtrA{C}(\rho^{A C}) =
\partialtrA{B C}(\rho^{A B C})$.

A similar formalism can be used to denote conditional quantum states.
Let $\rho^{A B}$ be a density operator on $\cH_A \otimes \cH_B$ and
let $\fZ=\{F_z\}_{z \in \cZ}$ be a POVM on $\cH_B$. Then
$\rho\AcondB{A}{\fZ=z}$ denotes the quantum state on $\cH_A$
conditioned on the event that the outcome of the measurement $\fZ$
applied to the subsystem $\cH_B$ equals $z \in \cZ$, i.e.,
\[
  \rho\AcondB{A}{\fZ=z}
:=
  \frac{1}{\tr((\id_A \otimes F_z) \rho\partAB{A}{B})} 
    \partialtrA{B}\bigl((\id_A \otimes F_z) \rho\partAB{A}{B}\bigr) \ .
\]
The notation can be extended to density operators over three and more
subsystems in an obvious way. Note that, since the partial trace and
the operation of conditioning on a measurement result commute, this is
compatible with our notation for partial states. For instance, if
$\rho^{A B C}$ is a tripartite density operator, then the conditional
states $\rho\AcondB{A}{\fZ=z}$ and $\rho\AcondB{A C}{\fZ=z}$ are
related by the partial trace, i.e., $\rho\AcondB{A}{\fZ=z} =
\partialtrA{C}(\rho\AcondB{A C}{\fZ=z})$.

We will use a similar formalism to denote the probability
distributions resulting from measurements of quantum states. Let
$\rho^A$ be a density operator and let $\fY=\{E_y\}_{y\in\cY}$ be a POVM on $\cH_A$.
Then $\rho\matone{A}{\fY}$ denotes the distribution of the outcome of
the measurement $\fY$ applied to $\rho^A$, i.e.,
\[
  \rho\matone{A}{\fY}(z) = \tr(E_y \rho) \ , \qquad \text{for all $y \in \cY$.}
\]
This can easily be generalized to product systems. For example, if
$\fY$ and $\fZ$ are POVMs on $\cH_A$ and $\cH_B$, respectively, then
$\rho\mresultAB{A}{B}{\fY}{\fZ}$ is the probability distribution of
the outcome of the product measurement $\fY \otimes \fZ$ applied to
$\rho^{A B}$.

Note that the operation of taking the partial trace of a density
operator has a classical analogue, namely taking the marginal
distribution.  Similarly, the operation of conditioning a quantum
state on a measurement result corresponds to conditioning a
probability distribution on the value of a random variable. Our
formalism is consistent with respect to these operations in the sense
that the following diagram commutes.

\[
\newcommand*{\lt}[1]{\makebox[8em]{\scriptsize #1}}
\newcommand*{\ltm}[1]{\makebox{\scriptsize #1}}
\begin{CD}
{\rho\AcondB{A}{\fZ=z}} @<{\lt{cond.}} 
<< {\rho^{A B}} @> {\lt{trace}} 
>> {\rho^A} \\
 @ V{\ltm{meas.}} VV @ V{\ltm{meas.}} VV @ V{\ltm{meas.}} VV \\
{\rho\mresultAcondB{A}{\fY}{\fZ=z}} @<{\lt{cond.}} 
<< {\rho\mresultAB{A}{B}{\fY}{\fZ}} @> {\lt{trace}} 
>> {\rho\matone{A}{\fY}} 
\end{CD}
\] 

Let $P$ be a probability distribution on $\cZ$ and let, for each $z
\in \cZ$, $\rho_z$ be a density operator on $\cH$. We will often write
the weighted sum of $\rho_z$ as an expectation value, i.e.,
\[
  \ExpE_{z \leftarrow P}[\rho_z] := \sum_{z \in \cZ} P(z) \rho_z \ .
\]
If the probability distribution is clear from the context, we only write
$\Exp_z[\rho_z]$.
For example, using our formalism, we have 
\begin{equation} \label{eq:expcond}
  \ExpE_{z \leftarrow \rho\mresultA{B}{\fZ}}[\rho\AcondB{A}{\fZ=z}]
= 
  \rho\partA{A} \ ,
\end{equation}
for any bipartite quantum state $\rho^{A B}$ on $\cH_A \otimes \cH_B$.
This is a simple reformulation of the fact that the partial state
$\rho^A$ on $\cH_A$ does not change when a measurement is applied to
the subsystem $\cH_B$.

\subsection{Distance measures}
Let $\probdistr(\cZ)$ be the set of probability distributions on the
set $\cZ$. The \emph{variational distance} between two probability
distributions $P, Q \in \probdistr(\cZ)$ is defined by
\[
  \cdist(P,Q) := \frac{1}{2} \sum_{z \in \cZ} |P(z) - Q(z)| \ .
\]
The variational distance is a metric on $\probdistr(\cZ)$. In particular,
$\cdist(P, Q) = 0$ if and only if $P=Q$, $\cdist$ is symmetric,
$\cdist(P, Q) = \cdist(Q,P)$, and the triangle inequality holds,
$\cdist(P, R) \leq \cdist(P, Q) + \cdist(Q, R)$. For two bipartite
distributions $P_{X Y}$ and $P_{X' Y'}$, the variational distance
cannot increase when taking the marginals,
\begin{equation} \label{eq:cdistpartialtr} 
  \cdist(P_X, P_{X'}) 
\leq 
  \cdist(P_{X Y}, P_{X' Y'}) \ .
\end{equation}
If $P_{X Y}$ and $P_{X' Y'}$ have the same marginals $P_{X} = P_{X'}$,
their distance can be expressed as the expectation value of the
distance between their conditional probability distributions,
\begin{equation} \label{eq:classicaldistanceidentity}
  \cdist(P_{XY},P_{X' Y'}) 
= 
  \ExpE_{x \leftarrow P_X}\bigl[\cdist(P_{Y|X=x},P_{Y'|X'=x})\bigr]
:=
  \sum_{x} P_X(x) \cdist(P_{Y|X=x},P_{Y'|X'=x}) \ .
\end{equation}
In particular, if the distributions have product form,
\begin{equation} \label{eq:distprod}
  \cdist(P_X \times P_Y, P_X \times P_{Y'}) = \cdist(P_Y, P_{Y'}) \ .
\end{equation}

A similar distance measure can be defined on the set $\Herm(\cH)$ of
hermitian operators on $\cH$. The \emph{trace distance} between two
operators $U, V \in \Herm(\cH)$ is defined by
\[
  \qdist(U,V) := \frac{1}{2} \tr \bigl| U-V \bigr| \ .
\]
Many properties of the variational distance also hold for the trace
distance. In particular, the trace distance is a metric on
$\Herm(\cH)$. Moreover, similarly to~\eqref{eq:cdistpartialtr}, the
trace distance cannot increase when taking the partial trace, i.e.,
for $U, V \in \Herm(\cH_A \otimes \cH_B)$,
\begin{equation} \label{eq:tracedistpartialtr} 
  \qdist\bigl(\partialtrA{B}(U), \partialtrA{B}(V) \bigr) 
\leq 
  \qdist(U, V) \ .
\end{equation}
We will also use the strong convexity of the trace distance, i.e., for
$U, U', V, V' \in \Herm(\cH)$ and $p, q \in [0,1]$ with $p+q=1$,
\[
  \qdist(p U + q U', p V + q V')
\leq
  p \qdist(U,V) + q \qdist(U',V') \ .
\]

The following lemma gives a simple expression for the trace distance
between two product operators $U \otimes V$ and $U' \otimes V$ with a
common factor $V$.

\begin{lemma}\label{lem:tracetensor}
  Let $U, U' \in \Herm(\cH_A)$ and $V \in \Herm(\cH_B)$. Then
  \[
  \qdist(U \otimes V, U' \otimes V) = \qdist(U, U') \cdot \tr(|V|).
  \]
\end{lemma}

\begin{proof}
  By definition,
  \begin{equation}\label{tracedistanceqprop}
  \qdist(U \otimes V, U' \otimes V) = \tr{|(U-U') \otimes V|}.
  \end{equation}
  We use the following general fact, which can be verified easily
  using the appropriate definitions. Let
  $f:\mathbb{C}\rightarrow\mathbb{C}$ a function satisfying
  \[
  f(a\cdot b)=f(a)\cdot f(b)\qquad\textrm{for all }a,b\in\mathbb{C}.
  \]
  Then
  \[
  f(A\otimes B)=f(A)\otimes f(B)
  \]
  for all $A,B \in \Herm(\cH)$. Applying this to
  equation~(\ref{tracedistanceqprop}) yields
  \[
  |(U-U')\otimes V|=|(U-U')|\otimes |V| \ .
  \]
  The assertion then follows from the identity 
  $\tr(A\otimes B)=\tr(A)\cdot\tr(B)$.
\end{proof}

As an immediate consequence of Lemma~\ref{lem:tracetensor}, we obtain
the equation
\begin{equation}\label{eq:qdistprod}
\qdist(\rho\otimes\sigma,\rho'\otimes\sigma)=\qdist(\rho,\rho')\ ,\
\end{equation}
for states $\rho,\rho'\in\states(\cH_A)$ and
$\sigma\in\states(\cH_B)$, which is the quantum analogue of
(\ref{eq:distprod}).  The trace distance between two density operators
$\rho$ and $\sigma$ on $\cH$ corresponds to the variational distance
between the probability distributions $\rho_{\fZ}$ and $\sigma_{\fZ}$
of the outcomes of a measurement applied to $\rho$ and $\sigma$,
respectively, for an optimal POVM $\fZ$ on $\cH$, i.e.,
\begin{equation} \label{eq:measdist}
  \qdist(\rho, \sigma) 
= 
  \max_{\fZ}\ \! \cdist(\rho_{\fZ}, \sigma_{\fZ}) \ .
\end{equation}

\subsection{Symmetry and exchangeability}

\subsubsection{Symmetric probability distributions and symmetric
  functions}

Let $\bz = (z_1, \ldots, z_n) \in \cZ^n$ and $\bzp = (z'_1, \ldots,
z'_m) \in \cZ^m$ be tuples of elements from a set $\cZ$.  We denote by
$\conc{\bz}{\bzp}$ the $(n+m)$-tuple obtained by concatenating $\bz$
and $\bzp$, $\conc{\bz}{\bzp} := (z_1, \ldots, z_n, z'_1, \ldots,
z'_m)$.

The \emph{frequency distribution} $\freq{\bz}$ of an $n$-tuple $\bz =
(z_1, \ldots, z_n) \in \cZ^n$ is the function on $\cZ$ defined
by\footnote{We denote by $[n]$ the set of natural numbers between $1$
  and $n$, i.e., $[n]:=\{1, \ldots, n\}$.}
\[
\freq{\bz}(z) := \frac{1}{n} \bigl|\{i \in [n]: z_i = z\}\bigr| \ ,
\quad \text{for every $z \in \cZ$,}
\]
i.e., $\freq{\bz}(z)$ is the relative number of occurrences of the
symbol $z$ in $\bz$. Note that $\freq{\bz}$ is a probability
distribution on $\cZ$, $\freq{\bz} \in \probdistr(\cZ)$. 

A symmetric function $f$ on $\cZ^n$ is a function such that $f(\bz)$
is invariant under permutations of the entries in $\bz$. In
particular, the value $f(\bz)$ only depends on the frequency
distribution $\freq{\bz}$ of $\bz$. For a formal definition, let $S_n$
be the set of permutations on $[n]$, and let, for any $\pi \in S_n$,
$\pi_{\cZ}$ be the bijection on $\cZ^n$ defined by
\[
  \pi_\cZ : (z_1, \ldots, z_n) \longmapsto (z_{\pi(1)}, \ldots, z_{\pi(n)}) \ ,
  \qquad \text{for all $(z_1, \ldots, z_n) \in \cZ^n$}.
\]

\begin{definition}
  A function $f$ with domain $\cZ^n$ is called \emph{symmetric} if
  \[
    f = f \circ \pi_{\cZ} \ , \qquad \text{for all $\pi \in S_n$.}
  \]
\end{definition}

In particular, a probability distribution $P_{\bZ} \in
\probdistr(\cZ^n)$ on $\cZ^n$ is called \emph{symmetric} if $P_{\bZ}$
is a symmetric function.  The following lemma is an immediate
consequence of these definitions.

\begin{lemma} \label{lem:symfunc}
  Let $\bZ$ be an $n$-tuple of random variables over a set $\cZ$ such
  that $P_{\bZ}$ is symmetric, and let $Y$ be a random variable over
  $\cY$ defined by a channel $P_{Y|\bZ}$ such that for every
  $y\in\cY$, the function $\bz\mapsto P_{Y|\bZ=\bz}(y)$ is symmetric.
  Then, for every $y\in\cY$, the conditional probability distribution
  $P_{\bZ|Y=y}$ is symmetric.
\end{lemma}

In particular, if $f: \cZ^n \rightarrow \cY$ is a symmetric function,
then, for any $y \in \cY$, $P_{\bZ|f(\bZ) = y}$ is symmetric. An
example is the function mapping any $n$-tuple $\bz$ to the frequency
distribution $\freq{\bz}$, i.e., $P_{\bZ|\freq{\bZ}=\freqfix}$ is
symmetric.\footnote{$P_{\freq{\bZ}}$ denotes the probability
  distribution of $\freq{\bz}$, for $\bz$ randomly chosen according to
  $P_{\bZ}$, that is, $P_{\freq{\bZ}}(\freqfix) := \PrE_{\bz
    \leftarrow P_{\bZ}}[Q_{\bz} = \freqfix]$. } The following lemma is
an immediate consequence of this fact.

\begin{lemma} \label{lem:symfreq}
  Let $\bZ = (Z_1, \ldots, Z_n)$ be an $n$-tuple of random variables
  over a set $\cZ$ such that $P_\bZ$ is symmetric.  Then, for any
  $\freqfix \in \probdistr(\cZ)$,
  \[
    P_{Z_i|\freq{\bZ}=\freqfix} = \freqfix \ , 
      \quad \text{for every $i \in [n]$.}
  \]
\end{lemma}

\subsubsection{Symmetric and exchangeable density operators} 

For any permutation $\pi \in S_n$, let $\pi_\cH$ be the unique 
endomorphism on $\cH^{\otimes n}$ satisfying
\[
  \pi_{\cH} (\ket{\phi_1} \otimes \cdots \otimes \ket{\phi_n})
=
  \ket{\phi_{\pi(1)}} \otimes \cdots \otimes \ket{\phi_{\pi(n)}} \ ,
\qquad \text{for all $\ket{\phi_1}, \ldots, \ket{\phi_n} \in \cH$.}
\]
It is easy to verify that $\pi_{\cH} \in \Lin(\cH^{\otimes n})$ is
unitary. A density operator $\rho^{B_1 \cdots B_n} \in
\Lin(\cH^{\otimes n})$ is called \emph{symmetric} if $\rho^{B_1 \cdots
  B_n} = \pi_{\cH} \rho^{B_1 \cdots B_n} \pi_{\cH}^{\dagger}$ for
every $\pi\in S_n$. The following definition generalizes this concept
to include an additional system $\cH_A$.

\begin{definition} \label{def:sym}
  A density operator  $\rho^{A B_1 \cdots B_n} \in \states(\cH_A\otimes \cH^{\otimes n})$ is
  called \emph{symmetric relative to} $\cH_A$ if 
  \[
    \rho^{A B_1 \cdots B_n}
  =
    (\id_A \otimes \pi_{\cH}) \rho^{A B_1 \cdots B_n} (\id_A \otimes \pi_{\cH}^{\dagger})
  \ ,
    \qquad \text{for all $\pi \in S_n$.}
  \]
\end{definition}


Let $\rho^{A B_1 \cdots B_n} \in \states(\cH_A \otimes \cH^{\otimes
  n})$ be symmetric relative to $\cH_A$. Then, for any choice of $r$
distinct indices $i_1,\ldots, i_r \in [n]$, $r \in [n]$, the state
$\rho\partAB{A}{B_{i_1}\cdots B_{i_r}}$ is symmetric relative to
$\cH_A$. Moreover, since it only depends on the number $r$ of distinct
indices, we will write $\rho\partAB{A}{\tensspace{r}}$ instead of
$\rho\partAB{A}{B_{i_1}\cdots B_{i_r}}$.

Similarly, for every $\bz \in \cZ^s$, the density operator
$\rho\AcondB{A B_{i_1} \cdots B_{i_{n-s}}}{\fZ^{\otimes s}=\bz}$
obtained by conditioning a symmetric state $\rho^{A \notimes
  \tensspace{n}} = \rho^{A B_1 \cdots B_n}$ on the outcomes of a POVM
$\fZ = \{F_z\}_{z \in \cZ}$ applied to $s$ subsystems is independent
of the indices $i_1, \ldots, i_{n-s}$.  Additionally, as an immediate
consequence of Lemma~\ref{lem:condstatemeas} below, this conditional
state is still symmetric relative to $\cH_A$. We will thus use the
abbreviation $\rho\AcondB{A\tensspace{n-s}}{\fZ^{\notimesn s}=\bz}$.


\begin{lemma}\label{lem:condstatemeas}
  Let $\rho\partAB{A\notimes B}{\tensspace{n}}\in\states(\cH_A\otimes\cH_B \otimes \cH^{\otimes
  n})$ be symmetric relative
  to $\cH_A \otimes \cH_B$.  Then, for any POVM
  $\fY=\{E_y\}_{y\in\cY}$ on $\cH_A$ and every $y\in\cY$,
  $\rho\AcondB{B\notimes \tensspace{n}}{\fY=y}$ is symmetric relative
  to $\cH_B$.
\end{lemma}

\begin{proof}
  It suffices to show that, for any $\pi \in S_n$ and any $y \in \cY$,
  \[
      (\id_B\otimes \pi_{\cH})\partialtrA{A}
      \bigl((E_y \otimes \id_{B\notimes \tensspace{n}}) 
      \rho\partAB{A\notimes B}{\tensspace{n}} \bigr) 
      (\id_B\otimes\pi_{\cH}^\dagger)
  =
    \partialtrA{A}      
    \bigl((
      E_y \otimes \id_{B\notimes \tensspace{n}})  
      \rho\partAB{A\notimes B}{\tensspace{n}}  
    \bigr)
  \]
  where 
$\id_{B   \notimes \tensspace{n}}:= \id_B \otimes \id_{\cH}^{\otimes n}$.  We
  use the general identity
  \[
  U\partialtrA{A}(W)U'=\partialtrA{A}\bigl((\id_A\otimes
  U)W(\id_A\otimes U')\bigr)\] 
where $U,U'\in\Lin(\cH_B)$ and $W \in
  \Lin(\cH_A \otimes \cH_B)$, setting $U:=\id_B\otimes\pi_{\cH}$,
  $U':=\id_B\otimes\pi_{\cH}^\dagger$, and $W:=(E_y \otimes \id_{B
    \notimes \tensspace{n}}) \rho\partAB{A\notimes B}{\tensspace{n}}$.
  This leads to
  \begin{multline*}
    (\id_B\otimes \pi_{\cH})
     \partialtrA{A}
      \bigl((E_y \otimes \id_{B \notimes \tensspace{n}}) 
      \rho\partAB{A\notimes B}{\tensspace{n}} \bigr)
      (\id_B\otimes\pi_{\cH}^\dagger) \\
  \begin{aligned}
   = \partialtrA{A}\bigl((\id_{A\notimes B} \otimes \pi_{\cH}) 
       (E_y \otimes \id_{B\notimes \tensspace{n}})  
       \rho\partAB{A\notimes B}{\tensspace{n}} 
      (\id_{A\notimes B}\otimes \pi_{\cH}^\dagger) \bigr) \\
   =
    \partialtrA{A}\bigl((E_y \otimes \id_{B\notimes\tensspace{n}})
      (\id_{A\notimes B} \otimes \pi_{\cH})  
      \rho\partAB{A\notimes B}{\tensspace{n}}  
      (\id_{A\notimes B} \otimes \pi_{\cH}^\dagger) \bigr)     
  \end{aligned}
  \end{multline*}
  The assertion then follows since $\rho\partAB{A\notimes
    B}{\tensspace{n}}$ is symmetric relative to $\cH_A \otimes \cH_B$,
  that is,
  \[
      (\id_{A\notimes B} \otimes \pi_{\cH})
      \rho\partAB{A\notimes B}{\tensspace{n}}  
      (\id_{A\notimes B}\otimes \pi_{\cH}^\dagger) 
    =
      \rho\partAB{A\notimes B}{\tensspace{n}}\ . \qedhere
  \]
\end{proof}

The notation for symmetric quantum states introduced above can also be
used to denote symmetric probability distributions, e.g., resulting
from measuring a symmetric quantum state. Let, for instance, $\rho^{A
  \notimes \tensspace{n}} = \rho^{A \notimes B_1 \notimes \cdots
  \notimes B_n} \in \states(\cH_A \otimes \cH^{\otimes n})$ be
symmetric relative to $\cH_A$. Then, for POVMs $\fY$ and $\fZ$ on
$\cH_A$ and $\cH$, respectively, we write
$\rho\mresultAB{A}{\tensspace{n}}{\fY}{\fZ^{\notimesn n}}$ instead of
$\rho\matn{A}{B_1}{B_{n}}{\fY}{\fZ}{\fZ}$.

To formulate our theorems in a compact way, we will make use of the
notion of exchangeability. A symmetric 
density operator $\rho\partAB{}{\tensspace{n}}\in \states(\cH^{\otimes n})$
is said to be $m$-\emph{exchangeable}, for $m \geq n$, if there exists
a symmetric density operator $\sigma\partAB{}{\tensspace{m}} \in
\states(\cH^{\otimes m})$ such that $\rho\partAB{}{\tensspace{n}}$
is the partial trace of $\sigma\partAB{}{\tensspace{m}}$, i.e.,
$\rho\partAB{}{\tensspace{n}}=\sigma\partAB{}{\tensspace{n}}$. Similarly to
Definition~\ref{def:sym}, the definition of exchangeability can be
generalized to include an additional system $\cH_A$.

\begin{definition}
  A density operator $\rho\partAB{A}{\tensspace{n}}\in \states(\cH_A
  \otimes \cH^{\otimes n})$ which is symmetric relative to $\cH_A$ 
is called
  $m$-\emph{exchangeable relative to $\cH_A$} if there exists a
  density
  operator $\sigma\partAB{A}{\tensspace{m}} \in \Lin(\cH_A \otimes \cH^{\otimes m})$ such that $\sigma\partAB{A}{\tensspace{n}}$
  is symmetric relative to $\cH_A$ and
$\rho\partAB{A}{\tensspace{n}}=\sigma\partAB{A}{\tensspace{n}}$. We
refer to $\sigma\partAB{A}{\tensspace{m}}$ as an {\em extension} of $\rho\partAB{A}{\tensspace{n}}$.
\end{definition}
In the following, we will often use the same label for an extension of
a state. For example, we will denote an extension of
$\rho\partAB{A}{\tensspace{n}}$ to $m$ systems (for $m \geq n$) by
$\rho\partAB{A}{\tensspace{m}}$.

\subsection{Dual basis and quantum tomography} \label{quantumtomographysection}

\begin{definition} \label{def:dual}
  Let $\{e_i\}_{i \in \cN}$ be a family of vectors in a Hilbert space
  $\cH$. A family $\{f_i\}_{i \in \cN}$ is called a \emph{dual} of
  $\{e_i\}_{i \in \cN}$ if
  \[
    v = \sum_{i \in \cN} \spr{f_i}{v} e_i \ , \quad
    \text{for all $v \in \cH$,}
  \]
  where $\spr{f_i}{v}$ denotes the inner product between $f_i$ and
  $v$.
\end{definition}

Without proof, we state the following lemma known from linear algebra.

\begin{lemma} \label{lem:dualexistence}
  If $\{f_i\}_{i \in \cN}$ is a basis of $\cH$, then there exists a
  unique family $\{e_i\}_{i \in \cN}$ such that $\{f_i\}_{i \in \cN}$
  is a dual of $\{e_i\}_{i \in \cN}$.
\end{lemma}
Note that the set $\Lin(\cH)$ of endomorphisms on $\cH$ forms a
complex Hilbert space with inner product $(U,V) \mapsto \tr(U
V^\dagger)$. Similarly, the set $\Herm(\cH)$ of hermitian operators on
$\cH$ is a real Hilbert space with inner product $(U, V) \mapsto \tr(U
V)$. Hence, a family $\{F_z\}_{z \in \cZ}$ of elements from
$\Herm(\cH)$ is a dual of a family $\{F^*_z\}_{z \in \cZ}$ if
\begin{equation} \label{eq:hermdual}
  U = \sum_{z \in \cZ} \tr(F_z U) F^*_z \ , \quad
  \text{for all $U \in \Herm(\cH)$.}
\end{equation}
In particular, expression~\eqref{eq:hermdual} states that the operator
$U$ is fully determined by the values of the traces $\tr(F_z U)$.  The
following lemma generalizes this fact to product spaces.

\begin{lemma} \label{lem:tenssum}
  Let $\{F_z\}_{z \in \cZ}$ and $\{F^*_z\}_{z \in \cZ}$ be families of
  elements from $\Herm(\cH_B)$ such that $\{F_z\}_{z\in\cZ}$ is the dual of
  $\{F^*_z\}_{z\in\cZ}$. Then, for any $W \in \Herm(\cH_A \otimes \cH_B)$,
  \[
    W = \sum_{z \in \cZ} W_z \otimes F^*_{z} \ ,
  \]
  where $W_z := \partialtrA{B}((\id_A \otimes F_z) W)$, for all $z
  \in \cZ$.
\end{lemma}

\begin{proof}
  It is easy to verify that $\Herm(\cH_A \otimes \cH_B) = \Herm(\cH_A) \otimes
  \Herm(\cH_B)$. Hence  there exist operators $U_i \in \Herm(\cH_A)$ and $V_i
  \in \Herm(\cH_B)$  such that $W = \sum_i U_i
  \otimes V_i$.  By the linearity of the sum and the trace operator,
  it thus suffices to show that
  \[
    U \otimes V 
  = 
    \sum_{z \in \cZ} 
      \partialtrA{B}
        \bigl((\id_A \otimes F_z) (U \otimes V)\bigr) \otimes F^*_{z} 
  \]
  for any $U \in \Herm(\cH_A)$ and $V \in \Herm(\cH_B)$.  Since
  $\partialtrA{B}$ can be written as
  $\partialtrA{B}=\id_A\otimes\tr_B$ (where $\cH_A\otimes\mathbb{C}$
  is identified with $\cH_A$) we find
  \[
    \partialtrA{B}\bigl((\id_A \otimes F_z)  (U \otimes V)\bigr) 
  =
    \partialtrA{B}\bigl(U \otimes (F_z V)\bigr) 
  = 
    \tr(F_z  V) U \ ,
  \]
  and thus
  \[
    \sum_{z \in \cZ} 
      \partialtrA{B}
        \bigl((\id_A \otimes F_z)  (U \otimes V) \bigr) \otimes F^*_{z} 
  =
    U \otimes \left( \sum_{z \in \cZ} \tr(F_z V) F^*_z \right)
  \]
  The assertion then follows from~\eqref{eq:hermdual}.
\end{proof}

Let $\fZ = \{F_z\}_{z\in\cZ}$ be a POVM on $\cH_B$ and let $\{F^*_z\}_{z\in\cZ}$ be a
family of elements from $\Herm(\cH_B)$ such that $\{F_z\}_{z\in\cZ}$ is the dual
of $\{F^*_z\}_{z\in\cZ}$. 
 Definition~\ref{eq:hermdual} directly implies that 
any density operator $\rho^B$ on $\cH_B$ can be written as
\begin{equation}\label{eq:tomogr}
  \rho^B = \ExpE_{z \leftarrow \rho\mresultA{B}{\fZ}}\bigl[F^*_z\bigr] \ , 
\end{equation}
i.e., $\rho^B$ is fully determined by the probability distribution
$\rho\mresultA{B}{\fZ}$ of the outcomes when applying the
measurement $\fZ$ on $\rho^B$.  On the other hand, it is an immediate
consequence of 
Lemma~\ref{lem:tenssum} that, for any density operator
$\rho\partAB{A}{B}$ on $\cH_A \otimes \cH_B$,
\begin{equation} \label{eq:tom}
  \rho\partAB{A}{B} 
=
  \ExpE_{z \leftarrow \rho\mresultA{B}{\fZ}}
    \bigl[\rho\AcondB{A}{\fZ=z} \otimes F^*_{z}\bigr] \ .
\end{equation}

The above formulas are useful for quantum state tomography, that is, the
reconstruction of an unknown quantum state  $\rho$ given only the
statistics of measurement applied to identical copies of
$\rho$.  For example, it follows from (\ref{eq:tomogr}), the strong convexity of the trace distance
and Lemma~\ref{lem:tracetensor} that the estimate $\tilde{\rho}^B :=
\ExpE_{z \leftarrow\tilde{P}_{Z}}\bigl[F^*_z\bigr]$ is close to
$\rho^B$,
\begin{equation} \label{eq:tomacc}
  \qdist(\rho^B, \tilde{\rho}^B) 
\leq 
  \sum_{z \in \cZ} \bigl| P_Z(z) - \tilde{P}_Z(z) \bigr|\cdot \tr |F^*_z| \ .
\end{equation}
In particular, in order to obtain good estimates, one should choose a
POVM $\fZ$ such that the traces $\tr |F^*_z|$ are small.

\section{Informationally complete POVMs and duals \label{sec:icpovmandduals}}


\subsection{Symmetric informationally complete POVMs} \label{sec:restdim}

Intuitively, a POVM $\fZ = \{F_z\}_{z\in\cZ}$ is useful for tomography
if the distance between any two operators $F_z$ and $F_{z'}$ is large.
This is for instance the case for symmetric POVMs as defined below,
where the operators $F_z$ are symmetrically distributed over the space
of positive operators.

\begin{definition}
  Let $\cH$ be a $d$-dimensional Hilbert space. A \emph{symmetric}
  informationally complete POVM $\fZ = \{F_z\}_{z\in[d^2]}$ on $\cH$
  is an informationally complete POVM that consists of rank-one
  projectors
  \[
  F_z:=\frac{1}{d}\ket{\psi_z}\bra{\psi_z}\qquad\textrm{ for all }z\in [d^2]
  \]
  with the property that
  \[
  \tr(F_z F_{z'})=\theta_d\qquad\textrm{for all }z\neq z'.
  \]
  for some $\theta_d\in\mathbb{C}$.
\end{definition}

Analytic constructions of symmetric informationally complete POVMs are
known for dimensions $d=2,3,4,6,8$ (see, e.g., \cite{joseph,grassl}).
It can be shown that if a symmetric informationally complete POVM
exists in dimension $d$, then $\theta_d$ is a universal constant which
is independent of the particular symmetric informationally complete
POVM.  It equals
\[
\theta_d=\frac{1}{d^2(d+1)}.
\]

\begin{lemma} \label{lem:symmetricPOVMdual}
  Let $\fZ=\{F_z\}_{z\in [d^2]} \in \POVM{\cH}$ be a symmetric
  informationally complete POVM on a $d$-dimensional Hilbert space
  $\cH$.  Then there is a set of operators
  $\{F^*_z\}_{z\in[d^2]}\subset\Herm(\cH)$ that satisfies
\begin{enumerate}
\item $\fZ$ is a dual of $\{F^*_z\}_{z\in[d^2]}$.
  \label{beindualbasissymmetricp}
\item For every $z\in [d^2]$, the eigenvalues of $F^*_z$ and their
  multiplicities are
\begin{eqnarray*}
\begin{matrix}
\lambda_0 &:=&d\qquad & n_0 &:=& 1\\
\lambda_1 &:=&-1 & n_1 &:=& d-1.
\end{matrix}
\end{eqnarray*} \label{beindualbasissymmetrictwo}
\end{enumerate}
\end{lemma}

\begin{proof}
  Let us define $\alpha := d^2+d-1.$ It is straightforward to verify
  that the operators
  \begin{eqnarray*}
  F^*_z&:=&\alpha F_z-\sum_{z'\neq z} F_{z'}\qquad z\in [d^2]
  \end{eqnarray*}
  satisfy $\tr(F_z^*F_{z'})=\delta_{z{z'}}$, where $\delta_{zz'}$
  denotes the Kronecker-delta, which equals one if $z=z'$ and $0$
  otherwise.  This implies property~\ref{beindualbasissymmetricp}.  To
  obtain their eigenvalues, consider the matrix
  \[
  B:=F^*_z+\id =(\alpha+1)F_z.
  \]
  Because $F_z=\frac{1}{d}\ket{\psi_z}\bra{\psi_z}$, the eigenvalues
  of $B$ are $\frac{\alpha+1}{d}$ and $0$, occurring with
  multiplicities $1$ and $d-1$, respectively. Hence
  statement~\ref{beindualbasissymmetrictwo} follows.
\end{proof}

\subsection{A construction for arbitrary dimensions} \label{sec:gendim}

As mentioned in the previous section, symmetric informationally
complete POVMs are suitable for tomography. However, their existence
is only proven for certain dimensions. In this section, we will give a
construction of informationally complete POVMs for any dimension. It
is motivated by a general group-theoretic technique for finding such
POVMs (see e.g., \cite{ariano}).

Let $\cH$ be a $d$-dimensional Hilbert space and let
$\omega:=e^{\frac{2\pi i}{d}}$ be the $d$-th primitive root of unity.
Define the operators
\begin{equation*}
D_{j k}:=\omega^{\frac{j\odot k}{2}}\sum_{m\in\mathbb{Z}_d}\omega^{j m}\ket{k\oplus
m}\bra{m}\qquad\textrm{for all }(j,k)\in\mathbb{Z}_d\times\mathbb{Z}_d,
\end{equation*}
where $\oplus,\odot$ denotes addition and multiplication modulo $d$, respectively.
Furthermore, define
\begin{equation*}
c((j,k),(l,m)):=j m-kl\qquad\textrm{for all }(j,k),(l,m)\in\mathbb{Z}_d\times\mathbb{Z}_d.
\end{equation*}
We will use the simple identity
\begin{equation}\label{corthogonalrelation}
\sum_{\beta\in\mathbb{Z}_d\times\mathbb{Z}_d}\omega
^{c(\alpha,\beta)}=d^2\cdot \delta_{\alpha,0}\qquad\textrm{for all }\alpha\in\mathbb{Z}_d\times\mathbb{Z}_d.
\end{equation}
where  $\delta_{\alpha,\beta}$ denotes the Kronecker-delta, which
equals $1$ if $\alpha=\beta$ and $0$ otherwise. Note that this
identity directly follows from
\begin{equation}
\sum_{m\in\mathbb{Z}_d}\omega^{km}=d\cdot\delta_{k,0}\qquad\textrm{ for
all } k\in\mathbb{Z}_d.\label{simpleomegaadd}
\end{equation}

\begin{lemma}\label{CComputation}
The operators $\{D_\alpha\}_{\alpha\in\mathbb{Z}_d\times\mathbb{Z}_d}$
have the following properties.
\begin{enumerate}
\item\label{CComputationfirst} $D_\alpha$ is unitary for every
  $\alpha\in\mathbb{Z}_d\times\mathbb{Z}_d$.
\item\label{CComputationsecond} For every $\rho\in\states(\cH)$, $
  \sum_{\alpha\in \mathbb{Z}_d\times\mathbb{Z}_d}D_\alpha\rho
  D^\dagger_\alpha=d\cdot\id$.
\item \label{CComputationthird} $D_\alpha D_\beta
  D^\dagger_\alpha=\omega^{c(\alpha,\beta)}D_\beta$, for all
  $\alpha,\beta\in\mathbb{Z}_d\times\mathbb{Z}_d$.
\item \label{CComputationintermed} $D^\dagger _\alpha=D_{-\alpha}$,
  for all $\alpha\in\mathbb{Z}_d\times\mathbb{Z}_d$.
\item\label{CComputationlast} $\tr(D^\dagger_\alpha D_\beta)=d\cdot
  \delta_{\alpha,\beta}$.
\end{enumerate}
\end{lemma}
\begin{proof}
  These properties are well-known (see e.g., \cite{ariano}) and can also be
  verified by direct calculation. The proof is omitted here.
\end{proof}

\begin{lemma}
Define\label{statecomplete}
\begin{equation*}
\rho:=\frac{d}{d^2+1}\id+\frac{1}{2d(d^2+1)}\sum_{\beta\in\mathbb{Z}_d\times\mathbb{Z}_d}(D_\beta+D^\dagger_\beta).
\end{equation*}
Then $\rho$ is a state, i.e., $\rho\in\states(\cH)$.
\end{lemma}
\begin{proof}
To show that the operator $\rho$ is non-negative, we make use of the
following operators. Let
\begin{eqnarray*}
\ket{\psi^{\nu}_{km}}&:=&\frac{1}{\sqrt{2}}(\omega^{\frac{\nu}{2}}\ket{k\oplus
m}+\omega^{-\frac{\nu}{2}}\ket{m})\qquad\textrm{and }\\
\rho^{\nu}_{km}&:=&\ket{\psi^{\nu}_{km}}\bra{\psi^{\nu}_{km}}\qquad\textrm{for }\nu\in
\mathbb{R},k,m\in\mathbb{Z}_d,k\neq 0\\
\rho^{\nu}_{0m}&:=&(1+\frac{\omega^\nu+\omega^{-\nu}}{2})\ket{m}\bra{m}.
\end{eqnarray*}
Note that $\rho^{\nu}_{0m}$ is not normalized, but non-negative (it
equals $0$ if $(\nu\mod d)=\frac{d}{2}$).

With the definition
\begin{equation*}
\nu(j,k,m):=\frac{j\odot k}{2}+j m \ ,
\end{equation*}
it is easy to verify that
\begin{equation*}
\sum_{m}\rho^{\nu(j,k,m)} _{km}=\id+\frac{1}{2}(D_{j k}+D^\dagger_{j k}) \ ,
\end{equation*}
and hence
\begin{equation*}
\rho=\frac{1}{d(d^2+1)}\sum_{j,k}\sum_{m}\rho^{\nu(j,k,m)} _{km} \ .
\end{equation*}
This implies that $\rho$ is indeed positive.
Let us show that $\rho$ given by this expression is correctly
normalized.
Note that
\begin{eqnarray*}
\sum_{j,k,m}\tr\left(\rho^{\nu(j,k,m)} _{km}\right) &=&
\sum_{j,m}\sum_{k\neq 0}\tr\left(\rho^{\nu(j,k,m)}
  _{km}\right)+\sum_{j,m}\tr\left(\rho^{\nu(j,0,m)}
  _{0m}\right)\\
&=&d^2(d-1)+\sum_{j,m}\left(1+\frac{\omega^{\nu(j,0,m)}+\omega^{-\nu(j,0,m)}}{2}\right) \ .
\end{eqnarray*}
Because of
\begin{eqnarray*}
\sum_{j,m}\omega^{\nu(j,0,m)}&=&\sum_{j,m}\omega^{j m}\\
&=&d,
\end{eqnarray*}
where the second equation follows from (\ref{simpleomegaadd}),
we obtain
\begin{equation}
\sum_{j,k,m}\tr\left(\rho^{\nu(j,k,m)} _{km}\right) = d(d^2+1).
\end{equation}
This concludes the proof.
\end{proof}

\begin{lemma}
  Let
  \begin{equation*}
  \Delta_\alpha := \frac{1}{d^2+1}\id+\frac{1}{2d^2(d^2+1)} \sum_{\beta\in\mathbb{Z}_d\times\mathbb{Z}_d} (\omega^{c(\alpha,\beta)}D_\beta+\omega^{-c(\alpha,\beta)}D^\dagger_\beta).
  \end{equation*}
  Then $\fZ :=
  \{\Delta_\alpha\}_{\alpha\in\mathbb{Z}_d\times\mathbb{Z}_d}$ is a
  POVM on $\cH$.
\end{lemma}

\begin{proof}
  We show that $\Delta_\alpha:=\frac{1}{d}D_\alpha\rho
  D^\dagger_\alpha$ where $\rho\in\states(\cH)$ is the state given in
  Lemma~\ref{statecomplete}. The statement then follows from
  Lemma~\ref{CComputation}~\ref{CComputationfirst}
  and~\ref{CComputationsecond} because these imply that the operators
  are non-negative and resolve the identity.
  
  We claim that
  \begin{equation}\label{usefulccomp}
D_\alpha D^\dagger_\beta D^\dagger_\alpha =\omega^{-c(\alpha,\beta)} D^\dagger_\beta.
  \end{equation}
  This can be verified as follows using the fact that the operators
  are unitary, Lemma~\ref{CComputation}~\ref{CComputationthird}, and
  the identity $c(\alpha,\beta)=-c(\beta,\alpha)$:
  \begin{eqnarray*}
D_\alpha D^\dagger_\beta D^\dagger_\alpha &=&
D^\dagger_\beta D_\beta D_\alpha D^\dagger_\beta D^\dagger _\alpha\\
&=& D^\dagger_\beta \omega^{c(\beta,\alpha)}D_\alpha
D^\dagger_\alpha\\
&=&\omega^{-c(\alpha,\beta)} D^\dagger_\beta.
  \end{eqnarray*}
  By inserting the state $\rho$ given in Lemma~\ref{statecomplete}, we
  obtain
  \begin{eqnarray*}
\frac{1}{d}D_\alpha\rho D^\dagger_\alpha
&=& \frac{1}{d^2+1}\id
+\frac{1}{2d^2(d^2+1)}\sum_{\beta\in\mathbb{Z}_d\times\mathbb{Z}_d}
(D_\alpha D_\beta D^\dagger_\alpha+D_\alpha D^\dagger_\beta
D^\dagger_\alpha).
  \end{eqnarray*}
  Using Lemma~\ref{CComputation}~\ref{CComputationthird} again as well
  as the identity~\eqref{usefulccomp} establishes the fact that this
  equals $\Delta_\alpha$.
\end{proof}

To simplify the notation, let us introduce the hermitian operators
\begin{equation*}
\Lambda_\alpha:=\frac{1}{2d}\sum_{\beta\in\mathbb{Z}_d\times\mathbb{Z}_d}(\omega^{c(\alpha,\beta)}D_\beta+\omega^{-c(\alpha,\beta)}D^\dagger_\beta).
\end{equation*}
With these operators, the POVM in question has the simple form
\begin{equation}
\Delta_\alpha:=\frac{1}{d^2+1}\cdot\left(\id+\frac{1}{d}\cdot\Lambda_\alpha\right)\qquad\textrm{for
all }\alpha\in\mathbb{Z}_d\times\mathbb{Z}_d.\label{POVMsymmetricin}
\end{equation}
Let us first compute two useful identities concerning the trace of
these operators.

\begin{lemma}\label{tracesoflambdalemma}
  For all $\alpha,\beta\in\mathbb{Z}_d\times\mathbb{Z}_d$
  \begin{eqnarray*}
\tr(\Lambda_\alpha)&=&1.\\
\tr(\Lambda^\dagger_\alpha \Lambda_\beta)&=&d\cdot \delta_{\alpha,\beta}.
  \end{eqnarray*}
\end{lemma}

\begin{proof}
  As $D_{0}=\id$, we have from Lemma
  \ref{CComputation}~\ref{CComputationlast} the identity
  $\tr(D_\beta)=d\cdot\delta_{\beta,0}$.  With
  \begin{equation}\label{tracedaggereq}
    \tr(A^\dagger)=\overline{\tr(A)}
  \end{equation} this gives
  \begin{eqnarray*}
    \tr(\Lambda_\alpha)
  =
    \frac{1}{2} (\omega^{c(\alpha,0)}+\omega^{-c(\alpha,0)})
  =
    1 \ .
  \end{eqnarray*}  
  Because of
  \begin{eqnarray*}
\Lambda^\dagger_\alpha \Lambda_\beta &=&\frac{1}{4d^2}\left(\sum_{\gamma}
\omega^{-c(\alpha,\gamma)}D^\dagger_\gamma+\omega^{c(\alpha,\gamma)}D_\gamma
\right)\left(\sum_{\delta}
\omega^{c(\beta,\delta)}D_\delta+\omega^{-c(\beta,\delta)}D^\dagger_\delta
\right)
  \end{eqnarray*}
  and identity (\ref{tracedaggereq}) we have
  \begin{eqnarray*}
\tr(\Lambda^\dagger_\alpha \Lambda_\beta)
&=&\frac{1}{4d^2}\sum_{\gamma,\delta}\left(\omega^{c(\beta,\delta)-c(\alpha,\gamma)}\tr(D^\dagger_\gamma
D_\delta)+\omega^{c(\alpha,\gamma)+c(\beta,\delta)}\tr(D_\gamma
D_\delta)\right)\qquad +\quad h.c.,
  \end{eqnarray*}
  where $h.c.$ denotes the complex conjugate of the previous
  expression.  Applying
  Lemma~\ref{CComputation}~\ref{CComputationintermed}
  and~\ref{CComputationlast} gives
  \begin{equation*}
\tr(\Lambda^\dagger_\alpha \Lambda_\beta)
=\frac{1}{4d}\cdot \sum_{\gamma}
\left(\omega^{c(\beta,\gamma)-c(\alpha,\gamma)}+
\omega^{c(\alpha,\gamma)+c(\beta,-\gamma)}\right)\quad + \quad h.c.
  \end{equation*}
  But $c(\beta,-\gamma)=-c(\beta,\gamma)$ and
  $c(\beta,\gamma)-c(\alpha,\gamma)=c(\beta-\alpha,\gamma)$, hence we
  obtain
  \begin{equation*}
\tr(\Lambda^\dagger_\alpha \Lambda_\beta)
=\frac{1}{4d}\cdot
\sum_{\gamma}(\omega^{c(\beta-\alpha,\gamma)}+\omega^{c(\alpha-\beta,\gamma)})\quad+ \quad h.c.
  \end{equation*}
  and thus finally
  \begin{equation*}
\tr(\Lambda^\dagger_\alpha \Lambda_\beta)=d\cdot
\delta_{\alpha,\beta},
  \end{equation*}
  as a consequence of equation (\ref{corthogonalrelation}).
\end{proof}

\begin{lemma}
  The POVM $\fZ$ is a dual of the family of hermitian operators
  \begin{equation*}
  \Theta_\alpha :=-d\left(\id-\frac{d^2+1}{d}\cdot\Lambda_\alpha\right)\qquad\alpha\in\mathbb{Z}_d\times\mathbb{Z}_d.
  \end{equation*}
\end{lemma}

\begin{proof}  
  The operators $\Theta_\alpha$ are hermitian since, by definition,
  the operators $\Lambda_\alpha$ are hermitian.  The fact that $\fZ$
  is a dual of the family of operators $\Theta_\alpha$ follows from
  the representation (\ref{POVMsymmetricin}) of the POVM operators and
  Lemma~\ref{tracesoflambdalemma}.
\end{proof}

\begin{lemma}\label{grouptheoryPOVMlemma}
  For every $\alpha\in\mathbb{Z}_d\times\mathbb{Z}_d$,
  $\tr(|\Theta_\alpha|)\leq d\cdot \sqrt{d^4+d^2-1}$.
\end{lemma}

\begin{proof}
  Let $\lambda_1,\ldots,\lambda_d$ be the eigenvalues of
  $\Theta_\alpha$ (including multiplicities). Then
  \begin{eqnarray*}
\tr(|\Theta_\alpha|) &=&\sum_{i=1} ^d |\lambda_i|\\
&\leq &d^{\frac{1}{2}}\sqrt{\sum_{i=1} ^d |\lambda_i|^2}\\
&=&d^{\frac{1}{2}}\sqrt{\tr(\Theta^\dagger_\alpha \Theta_\alpha)}.
  \end{eqnarray*}
  But $\tr(\Theta^\dagger_\alpha \Theta_\alpha)=d (d^4+d^2-1)$ as can
  be computed directly using Lemma~\ref{tracesoflambdalemma}.
\end{proof}

\section{Analysis of symmetric probability distributions} \label{sec:symmetricprobdistr}

In this section, we derive a number of useful properties of symmetric
probability distributions. These results will later be applied to
probability distributions resulting from measurements of a symmetric
quantum state.

It is worth noting that our proof of the finite quantum de Finetti
representation does not rely on a classical de Finetti-style theorem
(as opposed to~\cite{cavesfuchsschack}). It is, however,
straightforward to obtain a de Finetti representation for (classical)
probability distributions based on the results presented in the
sequel.

\subsection{Estimating the frequency distribution of a subsequence} \label{sec:estimatesfreq}

Let $\conc{\bz}{\bzb}$ be the concatenation of an $n$-tuple $\bz$ and
a $k$-tuple $\bzb$ of elements from $\cZ$. We show that, if
$\conc{\bz}{\bzb}$ is randomly chosen according to a symmetric
probability distribution $P_{\conc{\bZ}{\bZb}}$, then the frequency
distribution $\freq{\bzb}$ of the sub-tuple $\bzb$ is a good estimate
for the frequency distribution $\freq{\bz}$ of the remaining
subsequence $\bz$.

We need the following simple relation between the distances of
frequency distributions of subsequences obtained from a sequence of
elements from $\cZ$. 

\begin{lemma} \label{lem:boundconvert}
  Let $\cZ$ be a set and let $\bz$ and $\bzb$ be elements of $\cZ^n$
  and $\cZ^k$, respectively. Then
  \[
    \cdist(\freq{\bz},\freq{\bzb})
  \leq
    \frac{n+k}{n} \cdist(\freq{\bzb},\freq{\conc{\bz}{\bzb}}) \ .
  \]
\end{lemma}

\begin{proof}
  By the definition of the frequency distribution,
  \[
    (n+k) \freq{\conc{\bz}{\bzb}} 
  = 
    n \freq{\bz} + k \freq{\bzb} \ .
  \]
  Hence, using the convexity of the variational distance,
  \[ 
    \cdist(\freq{\conc{\bz}{\bzb}},\freq{\bz})
  =
    \cdist\left(\frac{n}{n+k} \freq{\bz} + \frac{k}{n+k} \freq{\bzb},\freq{\bz}\right)
  \leq
    \frac{k}{n+k} \cdist(\freq{\bzb},\freq{\bz}) \ .
  \]
  The triangle inequality then leads to
  \[
    \cdist(\freq{\bzb},\freq{\bz})
  \leq
    \cdist(\freq{\bzb},\freq{\conc{\bz}{\bzb}})
  + \cdist(\freq{\conc{\bz}{\bzb}},\freq{\bz})
  \leq
    \cdist(\freq{\bzb},\freq{\conc{\bz}{\bzb}})
  + \frac{k}{n+k} \cdist(\freq{\bzb},\freq{\bz}) \ ,
  \]
  from which the assertion follows.
\end{proof}

We now show that $\freq{\bz}$ is close to $\freq{\bzb}$ by showing
that the expression on the r.h.s. of the inequality in
Lemma~\ref{lem:boundconvert} is small in expectation.

\begin{lemma} \label{lem:distguessexp}
  Let $\bZ$ be an $n$-tuple and $\bZb$ a $k$-tuple of random variables
  over a set $\cZ$ of size $|\cZ|=t$ such that $P_{\conc{\bZ}{\bZb}}$
  is symmetric. Then, for $\conc{\bz}{\bzb} \leftarrow
  P_{\conc{\bZ}{\bZb}}$,
  \[
    \Exp_{\conc{\bz}{\bzb}}
      \! [\cdist(\freq{\bzb},\freq{\conc{\bz}{\bzb}})]
  \leq
    \frac{1}{2} \sqrt{\frac{t}{k}}\ .
  \]
\end{lemma}

\begin{proof}
  Let $\bZ:=(Z_1,\ldots,Z_n)$ and
  $\bZb:=(\bar{Z}_1,\ldots,\bar{Z}_k)$.  It suffices to show that, for
  all probability distributions $\freqfix \in
  \probdistr(\cZ)$,
  \begin{equation} \label{eq:distguessexpmain}
    \ExpE_{\bzb \leftarrow P_{\bZb|\freq{\conc{\bZ}{\bZb}} = \freqfix}}
      [\cdist(\freq{\bzb},\freqfix)]
  \leq
    \frac{1}{2} \sqrt{\frac{t}{k}}
  \end{equation}
  The assertion then follows from 
  \[
    \ExpE_{\conc{\bz}{\bzb} \leftarrow P_{\conc{\bZ}{\bZb}}}
      [\cdist(\freq{\bzb},\freq{\conc{\bz}{\bzb}})]
  =
   \ExpE_{\freqfix \leftarrow P_{\freq{\conc{\bZ}{\bZb}}}}
\bigl[      \ExpE_{\bzb \leftarrow P_{\bZb|\freq{\conc{\bZ}{\bZb}} = \freqfix}}
        [\cdist(\freq{\bzb},\freqfix)]\bigr] \ .    
  \]
  Let thus $\freqfix \in \probdistr(\cZ)$ be fixed. For every $z \in \cZ$, let
  $\chi_z$ be the function on $\cZ$ defined by $\chi_z(z') = 1$ if $z'
  = z$ and $\chi_z(z') = 0$ otherwise. Then
  \begin{equation} \label{eq:dsum}
    \cdist(\freq{\bzb},\freqfix)
  =
    \frac{1}{2} \sum_{z \in \cZ} | d_z(\bzb) |
  \end{equation}
  where, for any $\bzb = (\bar{z}_1, \ldots, \bar{z}_k)\in\cZ^k$, 
  \[
    d_z(\bzb)
  :=
    \frac{1}{k} \sum_{i=1}^k \bigl(\chi_z(\bar{z}_i) - \freqfix(z)\bigr) \ .
  \]
  Using Jensen's inequality,
  \begin{equation} \label{eq:dmodbound}
    \ExpE_{\bzb \leftarrow P_{\bZb|\freq{\conc{\bZ}{\bZb}}=\freqfix}}\bigl[|d_z(\bzb)|\bigr]
  \leq
    \sqrt{\ExpE_{\bzb \leftarrow P_{\bZb|\freq{\conc{\bZ}{\bZb}}=\freqfix}}
            \bigl[d_z(\bzb)^2\bigr]}
  \end{equation}
  We claim that for any $i \neq j$,
  \begin{equation} \label{easytoseeone}
    \ExpE_{(\bar{z}_i,\bar{z}_j) 
      \leftarrow P_{(\bar{Z}_i,\bar{Z}_j)|\freq{\conc{\bZ}{\bZb}}=\freqfix}}
      \bigl[
        (\chi_z(\bar{z}_i) - \freqfix(z))(\chi_z(\bar{z}_j) - \freqfix(z))
      \bigr] 
  \leq 
    0 \ .
  \end{equation}
  To prove this identity, first note that the expectation value can
  be written in the form
  \[
  \ExpE_{\bar{z}_i\leftarrow P_{\bar{Z}_i|\freq{\conc{\bZ}{\bZb}}=\freqfix}}
      \Bigl[
        \ExpE_{\bar{z}_j\leftarrow 
          P_{\bar{Z}_j|\freq{\conc{\bZ}{\bZb}}=\freqfix,\bar{Z}_i=\bar{z}_i}}
         \bigl[
           (\chi_z(\bar{z}_i) - \freqfix(z))(\chi_z(\bar{z}_j) - \freqfix(z))
         \bigr]
      \Bigr].
  \]
  Since $0\leq q(z)\leq 1$, it suffices to show that
  \begin{equation}\label{easytosee}
  \ExpE_{\bar{z}_j
    \leftarrow P_{\bar{Z}_j|\freq{\conc{\bZ}{\bZb}}=\freqfix,\bar{Z}_i=\bar{z}_i}}
  \bigl[
    \chi_z(\bar{z}_j) - \freqfix(z)
  \bigr] 
    \begin{cases}
      \leq 0\qquad\textrm{if }\bar{z}_i=z\\
      \geq 0\qquad\textrm{otherwise}.
    \end{cases}
  \end{equation}
  Because of Lemma~\ref{lem:symfunc}, the probability distribution
  $P_{\bZb|\freq{\conc{\bZ}{\bZb}}=\freqfix,\bar{Z}_i=\bar{z}_i}$ is symmetric.
  Hence by the definition of the frequency distribution, we have
  \[
  \ExpE_{\bar{z}_j
    \leftarrow P_{\bar{Z}_j|\freq{\conc{\bZ}{\bZb}}=\freqfix,\bar{Z}_i=\bar{z}_i}}
    \bigl[
      \chi_z(\bar{z}_j) 
    \bigr]
  =
    P_{\bar{Z}_j|\freq{\conc{\bZ}{\bZb}}=\freqfix,\bar{Z}_i=\bar{z}_i}(z)
  =
    \begin{cases}
      \frac{n q(z)-1}{n-1} & \textrm{if }z=\bar{z}_i \\
      \frac{n q(z)}{n-1} & \textrm{otherwise}.
    \end{cases}
  \]
  This proves (\ref{easytosee}) and thus (\ref{easytoseeone}).
  
  Using (\ref{easytoseeone}) and Lemma~\ref{lem:symfreq}, we obtain
  \begin{equation} \label{eq:dsquarebound}
  \begin{split}
    \ExpE_{\bzb 
      \leftarrow P_{\bZb|\freq{\conc{\bZ}{\bZb}}=\freqfix}}
    \bigl[d_z(\bzb)^2\bigr]
  & \leq
    \frac{1}{k^2} \sum_{i=1}^k
      \ExpE_{\bar{z}_i \leftarrow P_{\bar{Z}_i|\freq{\conc{\bZ}{\bZb}}=\freqfix}}
        \bigl[(\chi_z(\bar{z}_i) - \freqfix(z))^2\bigr] \\
  & =
    \frac{1}{k^2}\sum_{i=1} ^k\left(\freqfix(z) - \freqfix(z)^2\right)
  \leq
    \frac{\freqfix(z)}{k} \ .
  \end{split}
  \end{equation}
  Combining~(\ref{eq:dsum}), (\ref{eq:dmodbound}), and
  (\ref{eq:dsquarebound}) leads to
  \[
      \ExpE_{\bzb \leftarrow P_{\bZb|\freq{\conc{\bZ}{\bZb}} = \freqfix}}
        [\cdist(\freq{\bzb},\freqfix)]
  =
    \frac{1}{2} \sum_{z \in \cZ} \sqrt{\frac{\freqfix(z)}{k}} \ .
  \]
  The bound~(\ref{eq:distguessexpmain}) then follows from Jensen's
  inequality, which concludes the proof.
\end{proof}

\subsection{The product structure of symmetric probability
  distributions } \label{sec:prodclass}

Let $\bz$ be an $n$-tuple over $\cZ$ randomly chosen according to a
probability distribution $P_{\bZ}$ and let $\freqfix \in
\probdistr(\cZ)$ be an estimate for the frequency distribution
$\freq{\bz}$ of $\bz$. To quantify the quality of this estimate, it is
convenient to introduce the abbreviation $\expdist{P_\bZ}{\freqfix}$
for the expected distance between the actual frequency distribution
$\freq{\bz}$ and the estimate $\freqfix$, that is
\[
  \expdist{P_\bZ}{\freqfix} 
:= 
  \ExpE_{\bz \leftarrow P_{\bZ}}[\cdist(\freq{\bz},\freqfix)] \ .
\]
The main result of Section~\ref{sec:estimatesfreq} can then be
rephrased as follows.

\begin{lemma} \label{lem:freqExpdistExp}
  Let $\bZ$ be an $n$-tuple and $\bZb$ a $k$-tuple of random variables
  over a set $\cZ$ of size $|\cZ|=t$, for $k \leq n$, such that
  $P_{\conc{\bZ}{\bZb}}$ is symmetric. Then, for $\bzb \leftarrow
  P_{\bZb}$,
  \[
    \Exp_{\bzb}[\expdist{P_{\bZ|\bZb=\bzb}}{\freq{\bzb}}]
  \leq
    \sqrt{\frac{t}{k}} \ .
  \]
\end{lemma}

\begin{proof}
  From Lemma~\ref{lem:distguessexp} and Lemma~\ref{lem:boundconvert},
  \[
    \ExpE_{\conc{\bz}{\bzb} \leftarrow P_{\conc{\bZ}{\bZb}}}
      [\cdist(\freq{\bz},\freq{\bzb})]
  \leq
    \sqrt{\frac{t}{k}}
  \]
  The assertion then follows from the definition of
  $\expdist{\cdot}{\cdot}$.
\end{proof}

Lemma~\ref{lem:condindep} establishes a connection between the
quantity $\expdist{P_\bZ}{\freqfix}$ and the product structure of
symmetric probability distributions.

\begin{lemma} \label{lem:condindep}
  Let $Y$ be a random variable over $\cY$ and let $\bZ = (Z_1, \ldots,
  Z_r)$ be an $r$-tuple of random variables over $\cZ$, such that the
  conditional probability distribution $P_{\bZ|Y=y}$ is symmetric for
  every $y \in \cY$.  Then, for all $\freqfix \in \probdistr(\cZ)$,
  \[
    \cdist(P_{Y Z_i}, P_Y \times \freqfix)
  \leq 
    \expdist{P_{\bZ}}{\freqfix} \ , \quad \text{for every $i \in [r]$.}
  \]
\end{lemma}

\begin{proof}
  Using the strong convexity of the variational distance, we have
  \begin{equation} \label{eq:ZWconv}
    \cdist(P_{Y Z_i},P_Y \times \freqfix)
 \leq
  \ExpE_{\freqfixp \leftarrow P_{\freq{\bZ}}}
    \bigl[ \cdist(P_{Y Z_i|\freq{\bZ}=\freqfixp},
         P_{Y|\freq{\bZ}=\freqfixp} \times \freqfix) \bigr] \ .
  \end{equation}  
  Since $P_{\bZ|Y=y}$ is symmetric, Lemma~\ref{lem:symfreq} implies
  $P_{Z_i|Y=y, \freq{\bZ}=\freqfixp} = \freqfixp$, for every $y \in
  \cY$.  Hence,
  \begin{equation} \label{eq:ZWmarg}
    \cdist(P_{Y Z_i|\freq{\bZ}=\freqfixp},
      P_{Y|\freq{\bZ}=\freqfixp} \times \freqfix)
  =
    \cdist(P_{Y|\freq{\bZ}=\freqfixp} \times \freqfixp,
      P_{Y|\freq{\bZ}=\freqfixp} \times \freqfix) 
  =
    \cdist(\freqfixp,\freqfix)
  \end{equation}
  where the last equality follows from~\eqref{eq:distprod}.
  Combining~(\ref{eq:ZWconv}) and (\ref{eq:ZWmarg}) leads to
  \[
    \cdist(P_{Y Z_i},P_{Y} \times \freqfix)
  \leq
    \ExpE_{\freqfixp \leftarrow P_{\freq{\bZ}}}
      \bigl[ \cdist(\freqfixp,\freqfix) \bigr]
  =
    \ExpE_{\bz \leftarrow P_{\bZ}}
      [\cdist(\freq{\bz},\freqfix)] 
  =
    \expdist{P_{\bZ}}{\freqfix} \ .
  \]
\end{proof}

\section{Analysis of symmetric quantum states} \label{sec:symstates}

The goal of this section is to derive results on symmetric quantum
states, based on the corresponding results of symmetric probability
distribution given in the previous section. In
Section~\ref{sec:mainquantumtheorem}, we first show how certain
properties of the measurement statistics imply structural properties
of the corresponding quantum states. Then, in
Section~\ref{sec:producstructuresym}, we combine these results with
those of~Section~\ref{sec:prodclass} in order to prove statements
about the structure of symmetric quantum states.

\subsection{Deducing the state structure from measurement
  results} \label{sec:mainquantumtheorem}

Consider a state of a bipartite system conditioned on a measurement on
one of the systems. We first prove an upper bound on the amount of
dependence between
this conditional state  and the measurement outcome.

\begin{lemma}\label{lemmaalloperators}
  Let $\rho\partAB{A}{B} \in \states(\cH_A\otimes\cH_B)$ and let $\fZ
  = \{F_z\}_{z \in \cZ}$ be a POVM on $\cH_B$. Then, for all $\freqfix
  \in \probdistr(\cZ)$ and $z \in \cZ$,
  \[
    \qdist(\rho\AcondB{A}{\fZ=z},\rho\partA{A})
  \leq
    \frac{2}{\rho\mresultA{B}{\fZ}(z)}
      \max_{\fY} 
        \cdist(\rho\mresultAB{A}{B}{\fY}{\fZ},
               \rho\mresultA{A}{\fY} \times \freqfix) \ ,
  \]
where the maximization is over all POVMs $\fY$ on $\cH_A$.
\end{lemma}

\begin{proof}
  Let $\fY:=\{E_y\}_{y\in\cY}$ be a POVM on $\cH_A$.  Then, from the triangle
  inequality, \eqref{eq:cdistpartialtr} and~\eqref{eq:distprod},
  \begin{equation} \label{eq:tritwo}
  \begin{split}
    \cdist(
      \rho\mresultAB{A}{B}{\fY}{\fZ},
      \rho\mresultA{A}{\fY}\times \rho\mresultA{B}{\fZ}
    )
  & \leq 
    \cdist(
      \rho\mresultAB{A}{B}{\fY}{\fZ},
      \rho\mresultA{A}{\fY} \times \freqfix)
    + \cdist(\rho\mresultA{A}{\fY} \times \freqfix,
      \rho\mresultA{A}{\fY} \times \rho\mresultA{B}{\fZ}) \\
  & = 
    \cdist(
      \rho\mresultAB{A}{B}{\fY}{\fZ},
      \rho\mresultA{A}{\fY} \times \freqfix)
      + \cdist(\freqfix,\rho\mresultA{B}{\fZ}) \\
  & \leq 
    2D
  \end{split}
 \end{equation}
  where 
  \[
    D 
  :=  
    \max_{\fY} 
      \cdist(\rho\mresultAB{A}{B}{\fY}{\fZ},
             \rho\mresultA{A}{\fY} \times \freqfix) \ ,
  \]  
  with the maximization ranging over all POVMs $\fY$ on $\cH_A$.
  Using equation~\eqref{eq:classicaldistanceidentity}, we have
  \[
    \rho\mresultA{B}{\fZ}(z) \cdot \cdist\bigl(
          \rho\mresultAcondB{A}{\fY}{\fZ=z},\rho\mresultA{A}{\fY}
        \bigr)
  \leq
    \ExpE_{z \leftarrow \rho\mresultA{B}{\fZ}}
      \bigl[
        \cdist\bigl(
          \rho\mresultAcondB{A}{\fY}{\fZ=z},\rho\mresultA{A}{\fY}
        \bigr)
      \bigr]
  =
    \cdist(
      \rho\mresultAB{A}{B}{\fY}{\fZ}, 
      \rho\mresultA{A}{\fY} \times \rho\mresultA{B}{\fZ}
    ) \ .
  \]
  With~(\ref{eq:tritwo}), it follows that
  \[
    \cdist\bigl(\rho\mresultAcondB{A}{\fY}{\fZ=z},\rho\mresultA{A}{\fY}\bigr)
  \leq 
    \frac{2 D}{\rho\mresultA{B}{\fZ}(z)}
  \]
  for every $z \in \cZ$ and every POVM $\fY$ on $\cH_A$. Therefore,
  using~\eqref{eq:measdist}, we conclude that
  \[
    \qdist(\rho\AcondB{A}{\fZ=z},\rho^A)
  =
    \max_{\fY}
      \cdist\bigl(\rho\mresultAcondB{A}{\fY}{\fZ=z},\rho\mresultA{A}{\fY}\bigr)
  \leq 
    \frac{2D}{\rho\mresultA{B}{\fZ}(z)} \ . \qedhere
  \]
\end{proof}

For a POVM $\fZ = \{F_z\}_{z\in\cZ}$ on a Hilbert space $\cH$, we
define the constants
\begin{eqnarray*}
  \Cone(\fZ) 
&:= &
  \min_{\{F^*_{z}\}_{z\in\cZ} }2\cdot\left( \sum_{z \in
      \cZ}\tr(|F^*_z|) \right)\\
\Ctwo(\fZ) &:=&\sqrt{|\cZ|}\cdot\Cone(\fZ).
\end{eqnarray*}
where the minimum ranges over all families $\{F^*_{z}\}_{z \in \cZ}$
of elements from $\Herm(\cH)$ such that $\fZ$ is the dual of
$\{F^*_{z}\}_{z \in \cZ}$. If no such family $\{F^*_z\}_{z\in \cZ}$
exists, we set $\Cone(\fZ):=\infty$.
  
Note that, due to Lemma~\ref{lem:dualexistence}, if $\fZ$ is an
informationally complete POVM, then $\Cdim_i(\fZ) < \infty$, for
$i=1,2$. By the interpretation given to the values $\tr(|F^*_z|)$ in
Section~\ref{quantumtomographysection}, the value $\Cone(\fZ)$ can be
seen as a measure for the accuracy by which an unknown state $\rho$
can be estimated from the statistics obtained by the measurement
$\fZ$.

For a $d$-dimensional Hilbert space $\cH$, let $\Cdim_i(d)$ be the
minimum of $C_i(\fZ)$, minimized over all POVMs $\fZ$ on $\cH$, that
is
\[
  \Cdim_i(d):=\min_{\fZ} C_i(\fZ)\ , \qquad\textrm{for }i=1,2\ .
\]

The following corollary is a direct consequence of of
Lemma~\ref{lem:symmetricPOVMdual} and
Lemma~\ref{grouptheoryPOVMlemma}.

\begin{corollary}\label{constantscorollary}
  Let $\cH$ be a $d$-dimensional Hilbert space.  Then
\begin{equation*}
\begin{split}
\Cdim_1(d)&\leq 2\sqrt{2}\cdot d^5 \ .\\
\Cdim_2(d)&\leq 2\sqrt{2}\cdot d^6 \ .
\end{split}
\end{equation*}
If symmetric informationally complete POVMs exist in dimension $d$,
then 
\begin{equation*}
\begin{split}
\Cdim_1(d)&\leq 2d^2(2d-1)\ .\\
\Cdim_2(d)&\leq 2d^3(2d-1)\ .
\end{split}
\end{equation*}

\end{corollary}

The main result of this section expresses the intuitive fact that a
bipartite state has product form if every bipartite measurement yields
a product distribution.

\begin{theorem}\label{thm:mainquantum}
  Let $\rho\partAB{A}{B} \in \states(\cH_A \otimes \cH_B)$ and let
  $\fZ$ be a POVM on $\cH_B$. Then, for any $q \in \probdistr(\cZ)$,
  \[
    \qdist(\rho\partAB{A}{B},\rho\partA{A}\otimes\rho\partA{B})
  \leq 
    \Cone(\fZ) \cdot \! \max_{\fY  }
      \cdist(\rho\mresultAB{A}{B}{\fY}{\fZ},
        \rho\mresultA{A}{\fY} \times q) \ .
  \]
In particular, if $\cH_B$ is $d$-dimensional, then 
\[
  \qdist(\rho\partAB{A}{B},\rho\partA{A}\otimes\rho\partA{B})
  \leq \Cdim_1(d)\cdot\max_{\fY ,\fZ }
      \cdist(\rho\mresultAB{A}{B}{\fY}{\fZ},
        \rho\mresultA{A}{\fY} \times  \rho\mresultA{B}{\fZ}) \ .
\]
(The maxima are taken over all POVMs $\fY$ on $\cH_A$ and $\fZ$ on
$\cH_B$, respectively.)
\end{theorem}

\begin{proof}
  Let $\{F^*_z\}_{z\in \cZ}$ be a family of elements from
  $\Herm(\cH_B)$ such that $\fZ=\{F_z\}_{z \in \cZ} $ is the dual of
  $\{F^*_z\}_{z\in \cZ}$.  Then,
  according to \eqref{eq:tom},
  \[
    \rho\partAB{A}{B} 
  = 
    \ExpE_{z \leftarrow \rho\mresultA{B}{\fZ}}
      \bigr[\rho\AcondB{A}{\fZ=z} \otimes F^*_z\bigr]
  \qquad \text{and} \qquad
    \rho^B = \ExpE_{z \leftarrow \rho\mresultA{B}{\fZ}}\bigl[F^*_z\bigr] \ .
  \] 
  Using these identities, the strong convexity of the trace distance,
  and Lemma~\ref{lem:tracetensor}, we obtain
  \[
  \begin{split}
    \qdist(\rho\partAB{A}{B},\rho^A\otimes\rho^B)
  & = 
    \qdist(  \ExpE_{z \leftarrow \rho\mresultA{B}{\fZ}}
               \bigl[\rho\AcondB{A}{\fZ=z} \otimes F^*_z\bigr],
             \ExpE_{z \leftarrow \rho\mresultA{B}{\fZ}}
               \bigl[\rho^A\otimes F^*_z\bigr]) \\
  & \leq 
     \ExpE_{z \leftarrow \rho\mresultA{B}{\fZ}}\bigl[
      \qdist(\rho\AcondB{A}{\fZ=z} \otimes F^*_z,\rho^A\otimes F^*_z)\bigr] \\
  & =
     \ExpE_{z \leftarrow \rho\mresultA{B}{\fZ}}
       \bigl[\qdist(\rho\AcondB{A}{\fZ=z},\rho^A) \cdot \tr(|F^*_z|)\bigr] \ .
  \end{split}  
  \]
  The assertion then follows from Lemma~\ref{lemmaalloperators} and
  the definition of $\Cone(\fZ)$.
\end{proof}

\subsection{The product structure of symmetric quantum
  states} \label{sec:producstructuresym}

We now combine the results of Section~\ref{sec:symmetricprobdistr} and
Section~\ref{sec:mainquantumtheorem}.  We start with a quantum
analogue of Lemma~\ref{lem:condindep}.

\begin{lemma} \label{lem:prodone}
  Let $\rho\partAB{A}{\tensspace{r}} \in \states(\cH_A \otimes
  \cH^{\otimes r})$ be symmetric relative to $\cH_A$ and let $\fZ =
  \{F_z\}_{z \in \cZ}$ be a POVM on $\cH$. Then, for any $\freqfix \in
  \probdistr(\cZ)$,
  \[
     \qdist(\rho\partAB{A}{\tensspace{1}},
           \rho\partA{A} \otimes \rho\partA{\tensspace{1}})
  \leq
   \Cone(\fZ) \, 
      \expdist{\rho\mresultA{\tensspace{r}}{\fZ^{\notimesn r}}}{\freqfix} 
   \]
  and
  \[
    \cdist(\rho\mresultA{\tensspace{1}}{\fZ},\freqfix)
  \leq 
    \expdist{\rho\mresultA{\tensspace{r}}{\fZ^{\notimesn r}}}{\freqfix} \ .
  \]
\end{lemma}

\begin{proof}
  Let $\fY$ be a POVM on $\cH_A$. Since, by
  Lemma~\ref{lem:condstatemeas}, $\rho\AcondB{\tensspace{r}}{\fY=y}$
  is symmetric, the probability distribution
  $\rho\mresultAcondB{\tensspace{r}}{\fZ^{\notimes r}}{\fY=y}$ is also
  symmetric, for all $y \in \cY$.  Lemma~\ref{lem:condindep} thus
  implies
  \begin{equation}\label{eq:genclassdist}
    \cdist(\rho\mresultAB{A}{\tensspace{1}}{\fY}{\fZ}
      , \rho\mresultA{A}{\fY} \times \freqfix)
  \leq 
    \expdist{\rho\mresultA{\tensspace{r}}{\fZ^{\notimesn r}}}{\freqfix} \ .
  \end{equation}
  The first assertion of the lemma then follows from
  Theorem~\ref{thm:mainquantum}.
  
  The second assertion of the lemma follows directly
  from~\eqref{eq:genclassdist} and property~\eqref{eq:cdistpartialtr}
  of the variational distance, i.e., $\cdist(
  \rho\mresultA{\tensspace{1}}{\fZ}, \freqfix) \leq \cdist(
  \rho\mresultAB{A}{\tensspace{1}}{\fY}{\fZ}, \rho\mresultA{A}{\fY}
  \times \freqfix)$.
\end{proof}

The next lemma shows  that symmetry imposes severe
constraints on the structure of quantum states. More precisely, if a
symmetric quantum state has product structure with respect to one of
its subsystems, this directly implies that the state has product
structure with respect to all its subsystems.
\begin{lemma}\label{lem:splitoffone}
  Let $\rho\partAB{A}{\tensspace{n}} \in \states(\cH_A \otimes
  \cH^{\otimes n})$ be symmetric relative to $\cH_A$. Then
  \[
    \qdist(\rho\partAB{A}{\tensspace{n}},
      \rho^A \otimes(\rho^{\tensspace{1}})^{\otimes n})
  \leq 
    n\cdot
    \qdist(\rho\partAB{A}{\tensspace{n}},
      \rho\partAB{A}{\tensspace{n-1}}\otimes\rho^{\tensspace{1}}) \ .
  \]
\end{lemma}

\begin{proof}
  Using the triangle inequality for the trace distance, we get
  \begin{equation} \label{eq:splittriangle}
    \qdist(\rho\partAB{A}{\tensspace{n}},
      \rho\partA{A}\otimes(\rho\partA{\tensspace{1}})^{\otimes n})
  \leq 
    \sum_{i=0} ^{n-1}
      \qdist(\rho\partAB{A}{\tensspace{n-i}}
                \otimes(\rho\partA{\tensspace{1}})^{\otimes i},
            \rho\partAB{A}{\tensspace{n-i-1}}
                \otimes(\rho\partA{\tensspace{1}})^{\otimes (i+1)}). \ 
 \end{equation}
 Since, by equation~\eqref{eq:qdistprod}, the trace distance does not
 change when tracing out the product state
 $(\rho\partA{\tensspace{1}})^{\otimes i}$, we have, for any $i \in
 [n]$,
  \begin{equation} \label{eq:splitequal}
    \qdist(\rho\partAB{A}{\tensspace{n-i}}
              \otimes(\rho\partA{\tensspace{1}})^{\otimes i},
          \rho\partAB{A}{\tensspace{n-i-1}}
              \otimes(\rho\partA{\tensspace{1}})^{\otimes (i+1)}) 
  =
    \qdist\bigl(\rho\partAB{A}{\tensspace{n-i}},
      \rho\partAB{A}{\tensspace{n-i-1}}
        \otimes\rho\partA{\tensspace{1}}\bigr) \ .
  \end{equation}
  Using the fact that the trace distance can only decrease when taking
  a partial trace (see~\eqref{eq:tracedistpartialtr}), we get
  \begin{equation} \label{eq:splittrace}
    \qdist(\rho\partAB{A}{\tensspace{n-i}},
           \rho\partAB{A}{\tensspace{n-i-1}}\otimes\rho\partA{\tensspace{1}})
  \leq 
    \qdist(\rho\partAB{A}{\tensspace{n}},
           \rho\partAB{A}{\tensspace{n-1}}\otimes\rho\partA{\tensspace{1}}) \ .
  \end{equation}
  Combining \eqref{eq:splitequal} with~\eqref{eq:splittrace} and
  inserting this into~\eqref{eq:splittriangle} concludes the proof.
\end{proof}

Combined with Lemma~\ref{lem:prodone}, we obtain an upper bound on the
distance between a symmetric state with $n$~subsystems and an $n$-fold product state.
\begin{corollary} \label{cor:prod}
  Let $\rho\partAB{A}{\tensspace{n+m}} \in \states(\cH_A \otimes
  \cH^{\otimes n+m})$ be symmetric relative to $\cH_A$ and let $\fZ =
  \{F_z\}_{z \in \cZ}$ be a POVM on $\cH$. Then, for any $q \in
  \probdistr(\cZ)$,
  \[
    \qdist(\rho\partAB{A}{\tensspace{n}},
           \rho\partA{A} \otimes (\rho\partA{\tensspace{1}})^{\otimes n})
  \leq 
     n \, \Cone(\fZ) \, 
      \expdist{\rho\mresultA{\tensspace{m+1}}{\fZ^{\notimesn m+1}}}
              {\freqfix} \ .
  \]
\end{corollary}

\begin{proof}
  Obviously, the density operator $\rho\partAB{A}{\tensspace{n+m}}$ is
  symmetric relative to $\cH_{A'} := \cH_A \otimes \cH^{\otimes n-1}$.
  Thus, with $r:=m+1$, we can write $\rho\partAB{A'}{\tensspace{r}}$
  instead of $\rho\partAB{A}{\tensspace{n+m}}$, and, similarly,
  $\rho^{A \notimes \tensspace{n}} = \rho^{A' \notimes \tensspace{1}}$
  and $\rho^{A \notimes \tensspace{n-1}} = \rho^{A'}$.  Hence, by
  Lemma~\ref{lem:prodone},
  \[
    \qdist(\rho^{A \notimes \tensspace{n}},
           \rho^{A \notimes \tensspace{n-1}} \otimes \rho^\tensspace{1})
  =
    \qdist(\rho^{A' \notimes \tensspace{1}},
           \rho^{A'} \otimes \rho^\tensspace{1})
  \leq 
    \Cone(\fZ) \, 
      \expdist{\rho\mresultA{\tensspace{r}}{\fZ^{\notimesn r}}}{\freqfix} \ .
  \]
  Applying Lemma~\ref{lem:splitoffone} concludes the proof.
\end{proof}

Finally, we give an upper bound on the expected value of the quantity
appearing on the r.h.s. in Corollary~\ref{cor:prod}. For this, we need
a quantum analogue of Lemma~\ref{lem:freqExpdistExp}.

\begin{lemma} \label{lem:dofreqExp}
  Let $\rho\partA{\tensspace{n+k}} \in \states(\cH^{\otimes n+k})$,
  for $k \leq n$, be symmetric and let $\fZ = \{F_z\}_{z \in \cZ}$ be
  a POVM on $\cH$ with $|\cZ|=t$.  Then, for
  $\bzb\leftarrow\rho\mresultA{\tensspace{k}}{\fZ^{\notimesn k}}$,
  \[
    \Exp_{\bzb}
      \bigl[
        \expdist{
          \rho\mresultAcondB{\tensspace{n}}{\fZ^{\notimes n}}
              {\fZ^{\notimes k}=\bzb}}{\freq{\bzb}}
      \bigr]
  \leq
    \sqrt{\frac{t}{k}} \ .
  \]
\end{lemma}

\begin{proof}
  The assertion follows directly from Lemma~\ref{lem:freqExpdistExp}
  and the fact that the probability distribution
  $\rho\mresultAB{\tensspace{n}}{\tensspace{k}}{\fZ^{\notimes
      n}}{\fZ^{\notimes k}}$ is symmetric.
\end{proof}


\section{Main results} \label{sec:mainresult}

Our main results immediately follow from the characterizations of
symmetric quantum states given in the preceding section.  Basically,
we consider the quantum state $\rho_{\bzb}\partA{A \notimes
  \tensspace{n}} := \rho\AcondB{A\notimes
  \tensspace{n}}{\fZ^{\notimesn k}=\bzb}$ on $\cH_A \otimes
\cH^{\otimes n}$ obtained by conditioning an exchangeable state
$\rho\partAB{A}{\tensspace{n+k}}$ on the outcomes $\bzb=(z_1, \ldots,
z_k)$ of a POVM $\fZ$ applied to $k$ subsystems $\cH$. We show that
$\rho_{\bzb}\partA{A \notimes \tensspace{n}}$ is close to a product
state $\rho_{\bzb}\partA{A} \otimes
(\rho_{\bzb}\partA{\tensspace{1}})^{\otimes n}$ where
$\rho_{\bzb}\partA{\tensspace{1}}$ is a density operator on a single
subsystem $\cH$.  Moreover, $\rho_{\bzb}\partA{\tensspace{1}}$ is
almost determined by the observed measurement statistics.




\begin{theorem} \label{thm:mainexp}
  Let $\rho\partAB{A}{\tensspace{n+k}} \in \states(\cH_A \otimes
  \cH^{\otimes n+k})$ be $(n+2k-1)$-exchangeable relative to $\cH_A$
  and let $\fZ = \{F_z\}_{z \in \cZ}$ be a POVM on $\cH$ with
  $|\cZ|=t$. For every $\bzb \in \cZ^k$, let $\rho_{\bzb}\partA{A
    \notimes \tensspace{n}} := \rho\AcondB{A\notimes
    \tensspace{n}}{\fZ^{\notimesn k}=\bzb}$. Then for
  $\bzb\leftarrow\rho\mresultA{\tensspace{k}}{\fZ^{\notimesn k}}$
   \[
\Exp_{\bzb} \Bigl[ \qdist\bigl( \rho_{\bzb}\partA{A\notimes
     \tensspace{n}}, \rho_{\bzb}\partA{A} \otimes (
   \rho_{\bzb}\partA{\tensspace{1}})^{\otimes n} \bigr) \Bigr] \leq
   \Ctwo(\fZ) \, \frac{n}{\sqrt{k}}
   \]
   where $\rho_{\bzb}\partA{\tensspace{1}} =
   \rho\AcondB{\tensspace{1}}{\fZ^{\notimesn k}=\bzb}$ is determined
   by
   \[
  \Exp_{\bzb}
     \Bigl[
       \cdist\bigl(\rho\mresultAcondB{\tensspace{1}}{\fZ}{\fZ^{\notimesn k}=\bzb},
              \freq{\bzb}\bigr)
     \Bigr]
   \leq
     \sqrt{\frac{t}{k}} \ .
   \]
\end{theorem}

\begin{proof}
  Let $m:=k-1$. According to the definition of exchangeability, there
  exists an extension $\rho\partAB{A}{\tensspace{n+k+m}} \in
  \states(\cH_A \otimes \cH^{\otimes n+k+m})$ of
  $\rho\partAB{A}{\tensspace{n+k}}$ which is symmetric relative to
  $\cH_A$.  For every $\bzb \in \cZ^k$, let
  $[\rho_{\bzb}]\partAB{A}{\tensspace{n+m}} := \rho\AcondB{A
    \notimes\tensspace{n+m}}{\fZ^{\notimesn k} = \bzb}$.
  
  By Lemma~\ref{lem:condstatemeas},
  $[\rho_{\bzb}]\partAB{A}{\tensspace{n+m}}$ is symmetric relative to
  $\cH_A$. Hence, from the second bound of Lemma~\ref{lem:prodone},
  with $r:=m+1$,
  \begin{equation} \label{eq:mainclbound}
    \cdist([\rho_\bzb]\mresultA{\tensspace{1}}{\fZ},
           \freq{\bzb})
  \leq
    \expdist{[\rho_{\bzb}]\mresultA{\tensspace{m+1}}{\fZ^{\notimesn m+1}}}
                {\freq{\bzb}} \   
  \end{equation}
  and, from Corollary~\ref{cor:prod},
  \begin{equation} \label{eq:maindistbound}
    \qdist([\rho_{\bzb}]\partAB{A}{\tensspace{n}},
           [\rho_{\bzb}]^A 
             \otimes ([\rho_{\bzb}]\partA{\tensspace{1}})^{\otimes n})
  \leq
    n \, \Cone(\fZ) \,
      \expdist{[\rho_{\bzb}]\mresultA{\tensspace{m+1}}{\fZ^{\notimesn m+1}}}
                {\freq{\bzb}} \ .
  \end{equation} 
  Since $k = m+1$, we can apply Lemma~\ref{lem:dofreqExp} to the state
  $\rho\partA{\tensspace{(m+1)+k}}$, i.e.,
  \[
    \Exp_{\bzb}
      \bigl[
        \expdist{[\rho_{\bzb}]\mresultA{\tensspace{m+1}}{\fZ^{\notimesn m+1}}}
                {\freq{\bzb}}
      \bigr] 
  \leq
    \sqrt{\frac{t}{k}} \ .
  \]
  The assertion follows by taking the expectation on both sides
  of~(\ref{eq:mainclbound}) and~(\ref{eq:maindistbound}),
  respectively.
\end{proof}

Using Markov's inequality, it is straightforward to turn
Theorem~\ref{thm:mainexp}, which expresses closeness in terms of
expected distance, into a statement providing a bound on the
probability that the distance is larger than a given value $\eps$.
However, by adapting the auxiliary results derived so far and using a
tail inequality by Hoeffding, we obtain a tighter bound. The
interested reader is referred to Appendix~\ref{sec:markov} for a
derivation of an alternative version of Theorem~\ref{thm:mainexp}.

Finally, as a simple corollary of Theorem~\ref{thm:mainexp}, we obtain
the following representation for finitely exchangeable quantum states.

\begin{corollary}[Finite Quantum de Finetti Representation]
  Let $\cH$ be a $d$-dimensional Hilbert space and let
  $\rho\partA{\tensspace{n}} \in \states(\cH^{\otimes n})$ be
  $(n+s)$-exchangeable. Then $\rho\partA{\tensspace{n}}$ is
  $\eps$-close to the convex hull of the set of $n$-fold product
  states $\{\sigma^{\otimes n} : \sigma \in \states(\cH) \}$, for
  $\eps = \sqrt{2} \, \Cdim_2(d) {n}/{\sqrt{s}}$.
\end{corollary}

\begin{proof}
  Let $k:=\lceil s/2 \rceil$ and let $\rho^{\tensspace{n+k}} \in
  \states(\cH^{\otimes n+k})$ be an $(n+2 k-1)$-exchangeable extension
  of $\rho\partA{\tensspace{n}}$.  Let $\fZ$ be an informationally
  complete POVM on $\cH$ and let, for any $\bzb \in \cZ^k$,
  $\rho_{\bzb}\partA{\tensspace{n}} :=
  \rho\AcondB{\tensspace{n}}{\fZ^{\notimesn k} = \bzb}$. We show that
  \begin{equation} \label{eq:dfmain}
    \qdist(\rho\partA{\tensspace{n}},
\Exp_{\bzb}      \bigl[
        (\rho_\bzb^{\tensspace{1}})^{\otimes n}\bigr
      ])
  \leq
    \sqrt{2} \, \Cdim_2(d) \frac{n}{\sqrt{s}} \ .
  \end{equation}
  where $\bzb\leftarrow\rho\mresultA{\tensspace{k}}{\fZ^{\notimesn
      k}}$ Since, by~\eqref{eq:expcond},
  \[
    \Exp_{\bzb}
      \bigl[\rho_{\bzb}^{\tensspace{n}}\bigr]
  = 
    \rho^{\tensspace{n}} \ ,
  \]
  we obtain, using the strong convexity of the trace distance
  \begin{equation} \label{eq:qdone}
    \qdist(\rho^\tensspace{n},
  \Exp_{\bzb}
          \bigl[(\rho_\bzb^{\tensspace{1}})^{\otimes n}\bigr])
  =
    \qdist(
  \Exp_{\bzb}
        \bigl[\rho_{\bzb}\partA{\tensspace{n}}\bigr],
  \Exp_{\bzb}
        \bigl[(\rho_\bzb\partA{\tensspace{1}})^{\otimes n}\bigr])
  \leq 
  \Exp_{\bzb}
    [
      \qdist(\rho_\bzb ^\tensspace{n},(\rho_\bzb^{\tensspace{1}})^{\otimes n})
    ] \ .
  \end{equation}
  Inequality~(\ref{eq:dfmain}) then follows directly from
  Theorem~\ref{thm:mainexp}.
\end{proof}

While this result is of interest in its on right, we point out that
taking the limit $s \to \infty$ directly gives the well-known quantum
de Finetti representation for \emph{infinitely} exchangeable quantum
states~\cite{HudMoo76}, thus providing yet another new (compare
\cite{cavesfuchsschack}) and conceptually simple proof.

\section{Acknowledgment}

We would like to thank Ueli Maurer and Thomas Holenstein for many
helpful discussions and their support.


\appendix

\section{A Markov-style version of Theorem~\ref{thm:mainexp}} \label{sec:markov}

While Theorem~\ref{thm:mainexp} provides a bound on the expected
distance between the conditional state $\rho_{\bzb}\partA{A\notimes
  \tensspace{n}}$ and the product state $\rho_{\bzb}\partA{A} \otimes
( \rho_{\bzb}\partA{\tensspace{1}})^{\otimes n}$,
Theorem~\ref{thm:mainmu} below gives an expression for the minimum
probability such that this distance is smaller than some given value.

\begin{theorem} \label{thm:mainmu}
  Let $\rho\partAB{A}{\tensspace{n+k}} \in \states(\cH_A \otimes
  \cH^{\otimes n+k})$ be $(n+2k-1)$-exchangeable relative to $\cH_A$
  and let $\fZ = \{F_z\}_{z \in \cZ}$ be a POVM on $\cH$ with
  $|\cZ|=t$. For every $\bzb \in \cZ^k$, let $\rho_{\bzb}\partA{A
    \notimes \tensspace{n}} := \rho\AcondB{A\notimes
    \tensspace{n}}{\fZ^{\notimesn k}=\bzb}$. Then, for all $\eps\geq
  0$, and for
  $\bzb\leftarrow\rho\mresultA{\tensspace{k}}{\fZ^{\notimesn k}}$,
  with probability at least $1- k e^{-{\eps^2}/{2} +1}$,
  \[
        \qdist\bigl(
          \rho_{\bzb}\partA{A\notimes \tensspace{n}},
         \rho_{\bzb}\partA{A}
          \otimes (
         \rho_{\bzb}\partA{\tensspace{1}})^{\otimes n}
      \bigr)
       < \frac{n}{\sqrt{k}} \Ctwo(\fZ) \, \eps
  \]  
  and
  \[
        \cdist\bigl(
          \rho\mresultAcondB{\tensspace{1}}{\fZ}{\fZ^{\notimesn k}=\bzb},
               \freq{\bzb}
        \bigr)
        < \sqrt{\frac{t}{k}} \eps \ .
  \]
\end{theorem}

The proof of this theorem essentially follows the lines of the proof
of Theorem~\ref{thm:mainexp}. The main difference is that
Lemma~\ref{lem:distguessexp} is replaced by a statement based on a
tail inequality due to Hoeffding~\cite{hoeffding} which applies to
hypergeometric distributions as defined below (for more details, see,
e.g., \cite{JaLuRu00}).

\begin{definition}
  The \emph{hypergeometric distribution} with parameters $n$, $m$, and
  $k$, denoted $\Hyp{n}{m}{k}$, is defined as the probability
  distribution of the random variable $S:=|\Gamma \cap [m]|$ where
  $\Gamma$ is a randomly chosen subset of $[n]$ of size $|\Gamma| =
  k$.
\end{definition}

\begin{lemma}[Hoeffding] \label{lem:hoeffding}
  Let $S$ be a random variable with $P_S = \Hyp{n}{m}{k}$. Then, for
  all $\ell \geq 0$,
  \[
    \PrS[S \leq k \frac{m}{n} - \ell] \leq e^{-\frac{\ell^2 n}{2 k m}} \ .
  \]
\end{lemma}

Lemma~\ref{lem:distguesshoeff} and Lemma~\ref{lem:freqExpdistmu} are
adapted versions of Lemma~\ref{lem:distguessexp} and
Lemma~\ref{lem:freqExpdistExp}, as proven in
Section~\ref{sec:estimatesfreq} and Section~\ref{sec:prodclass},
respectively.

\begin{lemma} \label{lem:distguesshoeff}
  Let $\bZ$ be an $n$-tuple and $\bZb$ a $k$-tuple of random variables
  over a set $\cZ$ of size $|\cZ|=t$ such that $P_{\conc{\bZ}{\bZb}}$
  is symmetric. Then, for any $\eps \geq 0$ and for $\conc{\bz}{\bzb}
  \leftarrow P_{\conc{\bZ}{\bZb}}$,
  \[
    \PrS_{\conc{\bz}{\bzb}}
      \bigl[\cdist(\freq{\bzb},\freq{\conc{\bzb}{\bz}}) \geq \eps\bigr] 
  \leq 
    t \, e^{-\frac{k \eps^2}{2 t}} \ .
  \]
\end{lemma}

\begin{proof}
  It suffices to show that, for all probability distributions
  $\freqfix \in \probdistr(\cZ)$,
  \begin{equation} \label{eq:fdistmain}
    \PrE_{\bzb \leftarrow P_{\bZb|\freq{\conc{\bZ}{\bZb}}=\freqfix}} 
      \bigl[\cdist(\freq{\bzb},\freqfix) \geq \eps\bigr] 
  \leq 
    t \, e^{- \frac{k \eps^2}{2 t}} \ .
  \end{equation}
  The assertion of the lemma then follows from
  \[
    \PrE_{(\bz,\bzb) \leftarrow P_{\conc{\bZ}{\bZb}}} 
      \bigl[\cdist(\freq{\bzb},\freq{\conc{\bz}{\bzb}}) \geq \eps\bigr] 
  =
    \ExpE_{\freqfix \leftarrow P_{\freq{\conc{\bZ}{\bZb}}}}
    \Bigl[
      \PrE_{\bzb \leftarrow P_{\bZb|\freq{\conc{\bZ}{\bZb}}=\freqfix}} 
      \!\!\!\!
      \bigl[\cdist(\freq{\bzb},\freqfix) \geq \eps\bigr]
    \Bigr] \ .
  \]
  Let thus $\freqfix \in \probdistr(\cZ)$ with
  $P_{\freq{\conc{\bZ}{\bZb}}}(\freqfix)>0$ be fixed. The variational
  distance between $\freq{\bzb}$ and $\freqfix$ can be written as
  \begin{equation} \label{eq:distsum}
    \cdist(\freq{\bzb},\freqfix)
  = 
    \sum_{z \in \cZ'} \max\bigl(\freqfix(z)-\freq{\bzb}(z), 0\bigr) \ .
  \end{equation}
  where $\cZ' := \{z \in \cZ: \freqfix(z) > 0\}$. It is easy to see
  that, for any $z \in \cZ$, the random variable $S_z:=k \cdot
  \freq{\bZb}(z)$, conditioned on the event $\freq{\conc{\bZ}{\bZb}} =
  \freqfix$, is distributed according to $\Hyp{n+k}{(n+k) \,
    \freqfix(z)}{k}$. For any $z \in \cZ'$, let $\eps_z := \eps
  \sqrt{\textfrac{\freqfix(z)}{t}}$. Lemma~\ref{lem:hoeffding} (with
  $\ell=k \eps_z$) then implies
  \[
    \PrE_{\bzb \leftarrow P_{\bZb|\freq{\conc{\bZ}{\bZb}} = \freqfix}} 
      \bigl[\freqfix(z)-\freq{\bzb}(z) \geq \eps_z\bigr] 
  \leq 
    e^{-\frac{k \eps_z^2}{2 \freqfix(z)}} 
  =
    e^{-\frac{k \eps^2}{2 t}} \ ,
  \]   
  and thus, using the union bound,
  \begin{equation} \label{eq:boundfa}
    \PrE_{\bzb \leftarrow P_{\bZb|\freq{\conc{\bZ}{\bZb}} = \freqfix}} 
      \bigl[\forall z \in \cZ' : \, 
        \freqfix(z)-\freq{\bZb}(z) < \eps \sqrt{\frac{\freqfix(z)}{t}}
      \bigr]
  \geq
    1-t \, e^{-\frac{k \eps^2}{2 t}} \ .
  \end{equation}
  Since, by Jensen's inequality, 
  \[
    \sum_{z \in \cZ'} \eps \sqrt{\frac{\freqfix(z)}{t}}
  \leq
    \eps \sqrt{\sum_{z \in \cZ'} \freqfix(z)} 
  = 
    \eps 
  \]
  the event in~\eqref{eq:boundfa} implies that the sum on the r.h.s.
  of~(\ref{eq:distsum}) is smaller than $\eps$, that is,
  $\cdist(\freq{\bzb},\freqfix) < \eps$.
  Inequality~(\ref{eq:fdistmain}) thus follows directly from the
  bound~\eqref{eq:boundfa}.
\end{proof}

\begin{lemma} \label{lem:freqExpdistmu}
  Let $\bZ$ be an $n$-tuple and $\bZb$ a $k$-tuple of random variables
  over a set $\cZ$ of size $|\cZ|=t$, for $k \leq n$, such that
  $P_{\conc{\bZ}{\bZb}}$ is symmetric. Then, for all $\eps \geq 0$ and
  for $\bzb \leftarrow P_{\bZb}$,
  \[
    \PrS_{\bzb}
      [\expdist{P_{\bZ|\bZb=\bzb}}{\freq{\bzb}} \geq \eps]
  \leq
    k \, e^{- \frac{k \eps^2}{2 t} + 1} \ .
  \]
\end{lemma}

\begin{proof}
  Let $\tau:=\frac{t}{k}$ and $\epsp:=\eps-\tau$. Note that, if $\eps
  \leq \tau$, the r.h.s. of the inequality in the lemma becomes larger
  than $1$ because $\eps \leq 1$, i.e., the assertion is trivially
  true. Thus we can assume that $\epsp>0$.  For all $\bzb \in \cZ^k$,
  let
  \[
    p_{\bzb} 
  := 
    \PrE_{\bz \leftarrow P_{\bZ|\bZb=\bzb}}
      [\cdist(\freq{\bz},\freq{\bzb}) \geq \epsp] \ .
  \]
  We then have, by Lemma~\ref{lem:distguesshoeff},
  \[
    \ExpE_{\bzb \leftarrow P_{\bZb}}[p_{\bzb}]
  =
    \PrE_{\conc{\bz}{\bzb} \leftarrow P_{\conc{\bZ}{\bZb}}}
      [\cdist(\freq{\bz},\freq{\bzb}) \geq \epsp]
  \leq
    t \, e^{-\frac{k \epsp^2}{2 t}} \ ,
  \]
  and, by Markov's inequality,
  \begin{equation} \label{eq:pzbound}
    \PrE_{\bzb \leftarrow P_{\bZb}}[p_\bzb \geq \tau] 
  \leq 
    \frac{t}{\tau} \, e^{-\frac{k \epsp^2}{2 t}} \ .
  \end{equation}
  With the definition
  \[
    \Lambda:=\{\bzb \in \cZ^k: p_{\bzb} < \tau\} \ ,
  \]
  the bound~(\ref{eq:pzbound}) can be rewritten as
  \[
    \PrE_{\bzb \leftarrow P_{\bZb}}[\bzb \notin \Lambda]
  \leq
    \frac{t}{\tau} \, e^{-\frac{k \epsp^2}{2 t}}
  \leq
    k \, e^{-\frac{k \eps^2}{2 t}+1} 
  \]
  where the second inequality follows from the observation that
  $\epsp^2 = (\eps-\tau)^2 \geq \eps^2 - 2\tau$ and $\tau =
  \frac{t}{k}$.
  
  It thus remains to be shown that, for any $\bzb \in \Lambda$, 
  \begin{equation} \label{eq:freqExpmumain}
    \expdist{P_{\bZ|\bZb=\bzb}}{\freq{\bzb}} 
  =
    \ExpE_{\bz \leftarrow P_{\bZ|\bZb=\bzb}}
      [\cdist(\freq{\bz},\freq{\bzb})] 
  < 
    \eps \ .
  \end{equation}
  Let thus $\bzb \in \Lambda$ be fixed. Then
  \[
    \ExpE_{\bz \leftarrow P_{\bZ|\bZb=\bzb}} \!
      [\cdist(\freq{\bz},\freq{\bzb})] 
  =
    \!\!\sum_{\substack{\bz \in \cZ^n \\ \cdist(\freq{\bz},\freq{\bzb}) < \epsp}} 
      \!\!\!\! P_{\bZ|\bZb=\bzb}(\bz)
      \cdist(\freq{\bz},\freq{\bzb}) \quad
    \!+ \!\!\sum_{\substack{\bz \in \cZ^n \\ \cdist(\freq{\bz},\freq{\bzb}) \geq \epsp}}
      \!\!\!\! P_{\bZ|\bZb=\bzb}(\bz)
      \cdist(\freq{\bz},\freq{\bzb}) 
  <
    \epsp + p_{\bzb} 
  \]
  from which the bound~(\ref{eq:freqExpmumain}) follows by the
  definition of $\Lambda$ and $\epsp + \tau = \eps$.
\end{proof}

Finally, using Lemma~\ref{lem:freqExpdistmu}, we directly obtain
Lemma~\ref{lem:dofreqmu} below which corresponds to
Lemma~\ref{lem:dofreqExp} of Section~\ref{sec:producstructuresym}.

\begin{lemma} \label{lem:dofreqmu}
  Let $\rho\partA{\tensspace{n+k}} \in \states(\cH^{\otimes n+k})$,
  for $k \leq n$, be symmetric and let $\fZ = \{F_z\}_{z \in \cZ}$ be
  a POVM on $\cH$ with $|\cZ|=t$. Then, for all $\eps \geq 0$ and for
  $\bzb \leftarrow \rho\mresultA{\tensspace{k}}{\fZ^{\notimesn k}}$,
  \[
    \PrS_{\bzb}
      \bigl[
        \expdist{
          \rho\mresultAcondB{\tensspace{n}}
            {\fZ^{\notimesn n}}{\fZ^{\notimesn k}=\bzb}}
          {\freq{\bzb}} \geq \eps
      \bigr]
  \leq
    k e^{-\frac{k\eps^2}{2 t} +1} \ .
  \]  
\end{lemma}

The proof of Theorem~\ref{thm:mainmu} is now similar to the proof of
Theorem~\ref{thm:mainexp}, where, instead of
Lemma~\ref{lem:dofreqExp}, Lemma~\ref{lem:dofreqmu} is used to bound
the r.h.s. of~(\ref{eq:mainclbound}) and~(\ref{eq:maindistbound}).

\end{document}